\documentclass[10pt,journal,compsoc]{IEEEtran}

\usepackage{amsmath}
\usepackage{amsfonts}

\usepackage[linesnumbered]{algorithm2e}
\usepackage{enumitem}

\usepackage{subcaption}

\usepackage{pgfplots}
\usepackage{pgfplotstable}
\usepackage{multirow}
\usepackage{booktabs}

\newcommand{\BigO}[1]{\mathcal{O}(#1)}
\DeclareMathOperator*{\argmax}{arg\,max}

\def\stripzero#1{\expandafter\stripzerohelp#1}
\def\stripzerohelp#1{\ifx 0#1\expandafter\stripzerohelp\else#1\fi}

\begin{document}
\onecolumn
© 2019 IEEE.  Personal use of this material is permitted.  Permission from IEEE must be obtained for all other uses, in any current or future media, including reprinting/republishing this material for advertising or promotional purposes, creating new collective works, for resale or redistribution to servers or lists, or reuse of any copyrighted component of this work in other works.

\twocolumn

\title{Efficient method for parallel computation of geodesic transformation on CPU}

\author{Danijel~\v{Z}laus,~\IEEEmembership{Member,~IEEE,}
        Domen~Mongus,~\IEEEmembership{Member,~IEEE}

\IEEEcompsocitemizethanks{\IEEEcompsocthanksitem D. \v{Z}laus and D. Mongus are with the Faculty of Electrical Engineering and
Computer Science, University of Maribor, 2000 Maribor, Slovenia\protect\\
E-mail: danijel.zlaus@um.si, domen.mongus@um.si}\thanks{The manuscript was submitted for review on April 1, 2019.}}

\markboth{IEEE TRANSACTIONS ON PARALLEL AND DISTRIBUTED SYSTEMS, MANUSCRIPT FOR REVIEW}{\v{Z}LAUS \MakeLowercase{\textit{et al.}}: EFFICIENT METHOD FOR PARALLEL COMPUTATION OF GEODESIC TRANSFORMATION ON CPU}

\IEEEtitleabstractindextext{\begin{abstract}
This paper introduces a fast Central Processing Unit (CPU) implementation of geodesic morphological operations using stream processing. In contrast to the current state-of-the-art, that focuses on achieving insensitivity to the filter sizes with efficient data structures, the proposed approach achieves efficient computation of long chains of elementary $3 \times 3$ filters using multicore and Single Instruction Multiple Data (SIMD) processing. In comparison to the related methods, up to $100$ times faster computation of common geodesic operators is achieved in this way, allowing for real-time processing (with over $30$ FPS) of up to $1500$ filters long chains, applied on $1024\times 1024$ images. In addition, the proposed approach outperformed GPGPU, and proved to be more efficient than the comparable streaming method for the computation of morphological erosions and dilations with window sizes up to $183\times 183$ in the case of using \texttt{char} and $27\times27$ when using \texttt{double} data types.
\end{abstract}

}

\maketitle

\IEEEdisplaynontitleabstractindextext

\IEEEpeerreviewmaketitle

\section{Introduction}
\label{sec:Introduction}

\IEEEPARstart{M}{athematical morphology} has been applied, to great effect, for the discovery of spatial structures within various types of data sources. These range from Light Detection and Ranging
(LiDAR) \cite{mongus2014ground}, satellite images \cite{tan2014hyperspectral} to medical sensors \cite{ly2015real} and other sensory systems \cite{shi2017terahertz, 
garduno2017new}. In particular, so-called geodesic operators have gained a lot of attention, due to their ability either to remove connected components or leave them intact. However, as they are computed by eroding or dilating a marker image iteratively, this typically results in long chains of sequential filters.
Therefore, they are computationally demanding, despite being based on simple elementary operations. 

To address the high computational cost of morphological filters, three techniques stand out in the literature. These focus on efficient computation of erosion and dilation by:

\begin{enumerate}
\item Decomposition of the structuring element into either smaller one-dimensional structuring elements or into chords, where both reduce the total amount of comparisons  \cite{shih1991decomposition, urbach2008efficient}.
\label{skill:decomposition}
\item Preprocessing the image to filter out inconsequential values and reduce the total number of comparisons per element \cite{van1992fast, gil2002efficient}. 
\label{skill:preprocessing}
\item The use of queue data structures to filter out inconsequential values on the fly without any preprocessing \cite{dokladal2011computationally, morard2011linear}.
\label{skill:queue}
\end{enumerate}

Speedups of the first two techniques are, typically, implemented on General Purpose Graphic Processing Units (GPGPUs), where image rows (or columns) are processed in parallel \cite{domanski2009parallel, thurley2012fast, moreaud2014fast}, while the last technique is also suitable for queue-based 1-D operators \cite{karas2015gpu}. However, decomposition and queue-based techniques are also well suited for Field-Programmable Gate Arrays (FPGAs) \cite{clienti2008system,deforges2013fast,bartovsky2014parallel, bartovsky2015morphological, torres2016fpga}. This is because they utilize spatial parallelism with pipeline computation, thus, allowing for multiple filters to be applied simultaneously during each clock cycle, while being displaced by some amount of pixels. However, the length of the synthesized pipeline has an upper limit, which depends on the amount of gate fabric available on a given FPGA chip. Calculating long chains of filters may, therefore, not always be possible to achieve during a single image iteration, thus, requiring multiple passes and the potential reconfiguration of the FPGA chip if the filters do not remain the same. Such an iterative and configurable FPGA architecture, called a Morphological Co-Processing Unit (MCPU), was proposed by \cite{bartovsky2015morphological}. It implements two types of configurable pipelines, one for geodesic operators using a $3\times3$ structuring element, and the other for more general purpose erosion (or dilation) with a configurable structuring element size, which is implemented using the pixel pump algorithm proposed by \cite{dokladal2011computationally}.

From an algorithmic perspective, decomposition of a structuring element into segments is the most common way to increase computational efficiency \cite{lam1998fast}. However, as was shown by \cite{xu1991decomposition}, this may only be applied to 8-connected convex structuring elements. On the other hand, when considering efficient computation of an arbitrary structuring element, decomposition into 1-D chords was proposed by \cite{urbach2008efficient}. Here, improved per pixel complexity is achieved by using batch processing of chords and lookup tables. Another popular approach to ensuring computational efficiency of morphological operators is by using preprocessing, in order to ensure that computations are not sensitive to the size of the structuring element.
The well-known van Herk/Gil-Werman (HGW) algorithm \cite{gil1993computing}, for example, achieves this using a min/max prefix-suffix buffer. However, these approaches are, in general, limited to symmetric (odd or even-sized) structuring elements. Recently, queue-based algorithms, such as \cite{dokladal2011computationally}, have become more established. Their key advantages are that they are not sensitive to the size of the structuring element, require only a single image iteration, and, as shown by \cite{bartovsky2015real}, are extendable using decomposition to a polygon structuring element. However, they fall short in comparison to specialized ones. Well-known libraries, such as OpenCV \cite{opencv_library} and SMIL \cite{faessel2014smil}, still use multiple implementations of decomposed erosion (dilation) filters, and, depending on the structuring element shape, the optimal method is selected during run-time. However, similar to GPGPU implementations, they do not consider parallel computation of multiple filters, but instead, focus on iterative filter computations using available multicore resources. Due to this, they fall short of the expected performance when used to compute geodesic operators, which are characterized by long filter chains. 

In order to address these challenges, the following contributions are provided within this paper:

\begin{itemize}
\item An SIMD erosion (dilation) $3\times3$ kernel that utilizes stream and in-place processing for improved time and memory demands when computing long filter chains.
\item A pipeline for parallel processing of multiple filters with implicit image data synchronization.
\item A run-time examination of CPU topology to determine an optimal thread pinning strategy which maximizes data locality in the CPU cache.
\item An extensive performance validation of common geodesic operators using the proposed method.
\end{itemize}

The rest of the paper is organized as follows: The definitions of the tested geodesic operators are given in Section 2. Section 3 introduces the proposed method and implementation details of the filter template for erosion. Results are presented in Section 4, while Section 5 concludes the paper.

\section{Geodesic operators}
\label{sec:theoretical_background}
\begin{figure*}[htb!]
\centering
\includegraphics{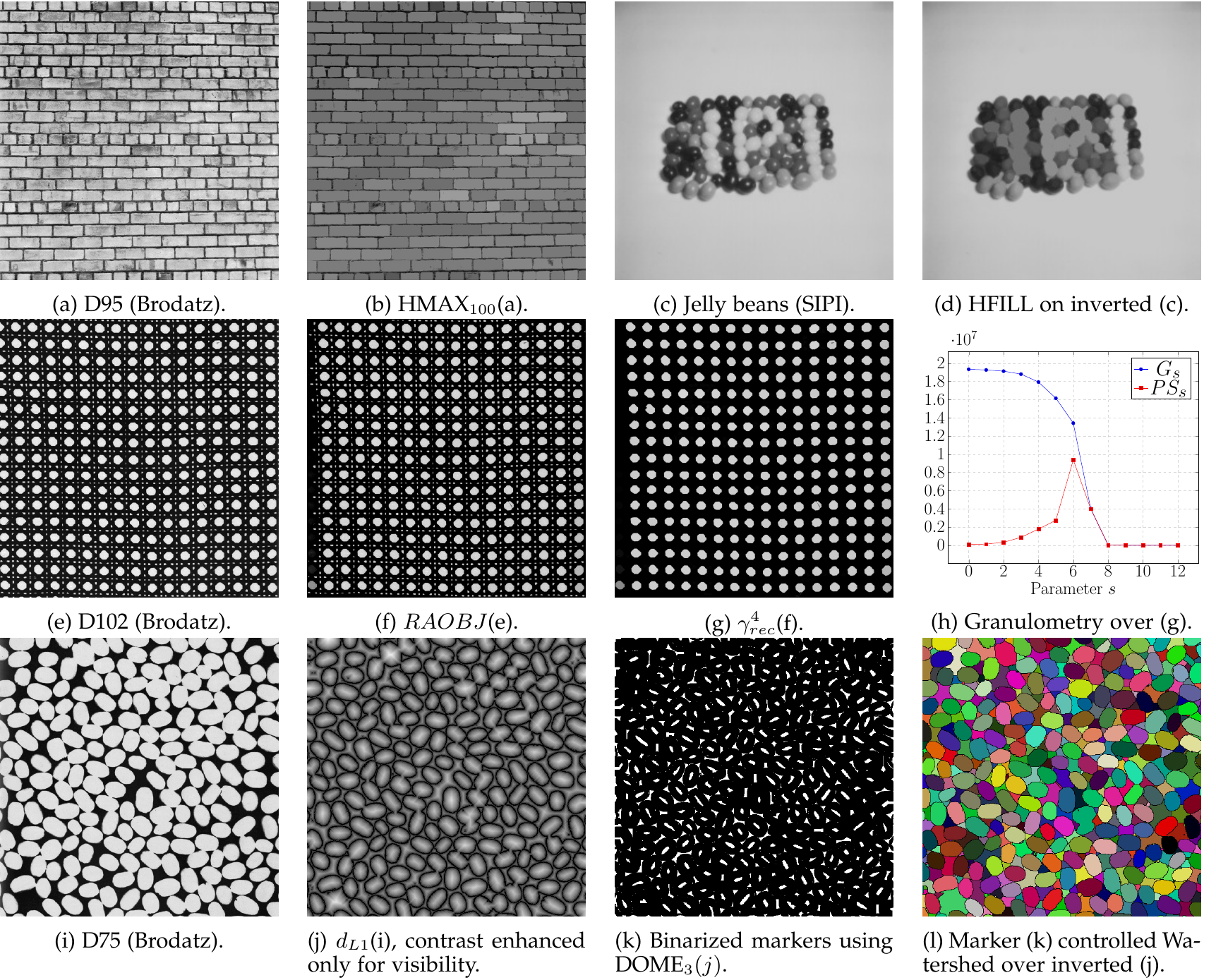}
\caption{Illustrative outputs of geodesic operators from Section \ref{sec:theoretical_background}.}
\label{fig:basic_geo_ops}
\end{figure*}
Let a gray-scale image $f$ be given as a mapping function $f: P \to \mathbb{R}$. In other words, $f$ defines an intensity value $f(p_i)$ to any given pixel position $p_i = (x_{i}, y_i)$, where $p_i \in P$. Note that $P$ is an ordered image domain given by a Cartesian product $P = \{0,1,\dots,X-1\} \times \{0,1,\dots,Y-1\}$, where $X$ is an image width and $Y$ is an image height. Let $w_s(p_i) = \{{x_{i} - s}, \dots, {x_{i} + s}\} \times \{{y_i - s}, \dots, {y_i + s}\}$, such that $w_s(p_i) \subseteq P$, be a square structuring element centered at $p_i$. Note, however, that $w_s$ is defined by its half size $s$ and, thus, it is limited to odd sizes. Accordingly, morphological erosion $\epsilon_s$ and dilation $\delta_s$ \cite{gonzales1987} are defined as:
\begin{flalign}
[\epsilon_s(f)](p_i) & = \min\limits_{q \in w_s(p_i)}{}f(q),\\
[\delta_s(f)](p_i) & = \max\limits_{q \in \overset{\vee}{w_s}(p_i)}{}f(q),
\end{flalign}
where $\overset{\vee}{w_s}$ is the reflected structuring element $w_s$.
By applying elementary operations sequentially, geodesic filters transform the marker image $f$ to the extent limited by the mask image $m$. A formal definition of elementary geodesic erosion $\epsilon_{1}^{m}$ is given as:
\begin{flalign}
\epsilon_{1}^{m}(f) & = \max(\epsilon_1(f), m).
\end{flalign}
Geodesic erosion $\epsilon_{s}^{m}(f)$ of size $s$ is then given as:
\begin{flalign}
\epsilon_{s}^{m}(f) & = \underbrace{\epsilon^{m}_{1} ( \epsilon^{m}_{1} ( \dots ( \epsilon^{m}_{1}}_{\text{$s$ times}}(f)))),
\label{eq:erosion_s}
\end{flalign}
while erosion by reconstruction $\epsilon_{rec}^{m}$ is defined as the convergence of sequentially concatenated geodesic erosions with regard to the mask image $m$:
\begin{flalign}\epsilon_{rec}^{m}(f) = \lim_{s \rightarrow \infty}
 \epsilon^{m}_{s}(f).
\label{eq:erosion_by_reconstruction}
\end{flalign}
Consequently, its dual geodesic dilation $\delta^{m}_{1}$ is defined as $\delta^{m}_{1}(f) = \min(\delta_1(f), m)$. Geodesic dilation $\delta_{s}^{m}(f)$ and geodesic dilation by reconstruction $\delta^{m}_{rec}(f)$ are then defined using elementary geodesic dilation $\delta^{m}_{1}(f)$, as follows from Eqs. \ref{eq:erosion_s} and \ref{eq:erosion_by_reconstruction}. 

These fundamental filters allow for the derivation of a family of geodesic operators. Some of the most popular ones are described next:
\begin{itemize}
\item H-maxima \cite{vincent1993morphological} suppress regional maxima and their variants whose contrast is lower than $h \in \mathbb{R}^+$:
\begin{flalign}
\text{HMAX}_h(f) = \delta^{f}_{rec}(f-h).
\label{eq:hmaxima}
\end{flalign}
\item Dome extraction \cite{vincent1993morphological, halkiotis2007automatic} is defined as the top-hat of H-maxima and extracts the removed regional maxima:
\begin{flalign}
\text{DOME}_{h}(f) = f - \text{HMAX}_h(f).
\label{eq:domeextract}
\end{flalign}
\item Hole filling \cite{soille1990automated, soille2007mathematical} removes all regional minima not attached to a marker image. In this context, $f$ and $m_\text{HFILL}$ are mask and marker images, respectively. Hole filling is defined as:
\begin{flalign}
\text{HFILL}(f) = \epsilon_{rec}^{f}(m_\text{HFILL}(f)),
\end{flalign}
while common marker $m_\text{HFILL}$ is defined using border pixels $B$ of mask $f$ as follows:
\begin{flalign}
 [m_\text{HFILL}(f)](p_i)= 
\begin{cases}
    f(p_i),\!&\!\text{if } p_i \in B \\
    \max\limits_{p\in P}{f(p)},\!&\!\text{otherwise.}
\end{cases}
\end{flalign}
\item Removal of objects attached to an image border is defined as:
\begin{flalign}
\text{RAOBJ}(f) &= f - \delta_{rec}^{f}(m_\text{RAOBJ}(f)), \label{eq:objremoval}
\end{flalign}
where common marker image $m_\text{RAOBJ}$ is derived from $f$ by:
\begin{flalign}
 [m_{RAOBJ}(f)](p_i)&= 
\begin{cases}
    f(p_i),\!&\!\text{if } p_i \in B \\
    \min\limits_{p\in P}{f(p)},\!&\!\text{otherwise.}
\end{cases}
\end{flalign}
\item Opening by reconstruction \cite{belgherbi2014morphological, pesaresi2001new} removes connected components smaller than $s$ from the image $f$. Its definition is given as:
\begin{flalign}
\gamma_{rec}^{s}(f) = \delta_{rec}^{f}(\epsilon_{s}(f)).
\label{eq:openbyrecon}
\end{flalign}
\item Quasi-distance transformation \cite{beucher2007numerical, hanbury2009morphological, brun2007restoration} maps the geodesic distance of each element to its corresponding largest residual value. Accordingly, it is defined based on a discontinuous function $d(f)$ that is given by:
\begin{flalign}
d(f) &= \argmax\limits_{s}[\epsilon_{s}(f) - \epsilon_{s+1}(f)].
\label{eq:quasi_eq1}
\end{flalign}
Iterative corrections are then applied on $d(f)$, until it converges to an 1-Lipschitz continuous function. This is achieved by:
\begin{flalign}
[\eta(f)](p_i)\!&=\!
\begin{cases}
    [\epsilon_1( f )](p_i)\!+\!1,\!&\text{if } [f\!-\!\epsilon_1( f )](p_i)\!>\!1 \\
    f(p_i),\!&\!\text{otherwise,}
\end{cases}
\label{eq:qdt_eq2}
\end{flalign}
which gives an $L_1$ norm quasi-distance transform $d_{L1}(f)$:
\begin{flalign}
d_{L1}(f) &= \lim_{s \rightarrow \infty}\underbrace{\eta ( \dots ( \eta }_{\text{$s$ times}}(d(f)))),
\end{flalign}
\item Granulometry \cite{matheron1975random} uses an increasing opening $\gamma_s$, or the dual operator closing $\varphi_s$, to produce a series of images $\{\gamma_0(f), \gamma_1(f), \dots \}$. Opening $\gamma_s$ is given as:	
\begin{flalign}
\gamma_s(f) = \delta_s(\epsilon_s(f)) &= \underbrace{\delta_1 ( \dots ( \delta_1 }_{\text{$s$ times}}( \underbrace{\epsilon_1 ( \dots ( \epsilon_1 }_{\text{$s$ times}}(f))))),
\label{eq:granulometry_func}
\end{flalign}
while the granulometric function $G_s(f)$ is given as: 
\begin{flalign}
G_s(f) &= \sum_{p \in P} [\gamma_s(f)](p).
\end{flalign}
The pattern spectrum, representing the size distribution of objects in the image, is given by the derivative of the granulometric function:
\begin{flalign}
PS_s(f) &= G_{s}(f) - G_{s+1}(f).
\end{flalign}
\item Alternating Sequential Filters (ASF) are constructed using sequentially concatenated openings $\gamma_s$ and closings $\varphi_s$, which are increasing in size. Closing $\varphi_s$ is given as:
\begin{flalign}
\varphi_s(f) = \epsilon_s(\delta_s(f)) = \underbrace{\epsilon_1 ( \dots ( \epsilon_1 }_{\text{$s$ times}}( \underbrace{\delta_1 ( \dots ( \delta_1 }_{\text{$s$ times}}(f))))),
\end{flalign}
while ASF up to size $s$ and starting with opening is given as:
\begin{flalign}
\text{ASF}_s(f) = \varphi_{s}(\gamma_{s}( \varphi_{s-1}(\gamma_{s-1}( \dots \varphi_1(\gamma_1(f)))))).
\end{flalign}
\end{itemize}

Illustrative outputs of the above geodesic operators are shown in Fig. \ref{fig:basic_geo_ops}. As is obvious from the definitions above, the main bottleneck of all these operators is the computation of sequentially concatenated morphological filters (see Section \ref{sec:results}). Its efficient implementation is, therefore, discussed next.

\section{Proposed method}
The proposed method enables efficient computation of geodesic filters composed from  sequentially concatenated elementary filters, such as $\epsilon^{m}_{1}$ and $\delta^{m}_{1}$. This is achieved by minimizing the required system memory bandwidth, reducing the number of mispredicted branches, and ensuring sequential image access. In this way, the advantages of the latency hiding facilities of modern CPU architectures are utilized fully, such as multi-level caches and hardware based memory prefetching. Furthermore, execution is speeded up further by utilizing available multicore resources. The proposed implementation consists of the following steps:
\begin{enumerate}
\item \textbf{decomposition of structuring element}, in order to ensure linear memory access,
\item \textbf{utilization of SIMD instructions}, with the purpose of speeding up the processing of decomposed filters,
\item \textbf{in-place processing}, with the objective to decrease total memory load,
\item \textbf{stream processing}, that ensures kernel results are obtained within a single top to bottom pass through the image $f$,
\item \textbf{efficient parallel processing}, of the sequential filter chain, 
\item \textbf{maximizing data locality}, by pinning threads based on run-time examination of the CPU cache topology, and
\item \textbf{implementation details of geodesic operators}, based on the proposed $3 \times 3$ kernel template.
\end{enumerate}
In the continuation these steps are described in detail, using the erosion of $f$ only as an example, while methodological extensions to dilation and data of higher dimension are obvious.

\subsection{Decomposition of structuring element}

As shown by \cite{lam1998fast}, the direct computation of morphological filters using structuring element $w_s(p_i)$ requires $\BigO{(2s+1)^2-1}$ comparisons per pixel position. It, therefore, follows that $8$ comparisons are required when considering $w_1(p_i)$.
One simple approach to improve this is by decomposing the structural element $w_1$ into its horizontal $w^x_{1}$ and vertical $w_1^y$ components. As shown by \cite{soille2013morphological}, this is achieved by:
\begin{flalign}
w^x_1(p_i) &=  \{x_{i} - 1, x_{i}, x_{i} + 1\} \times \{y_i\},\\
w^y_1(p_i) &= \{x_{i}\} \times \{y_i - 1, y_i, y_i + 1\},\\
\epsilon_{1}(f) &= \epsilon_1^x(\epsilon_1^y(f)) = \epsilon_1^y(\epsilon_1^x(f)).
\end{flalign}
This decreases the number of required comparisons per pixel to $4$. However, the computational efficiency of $\epsilon_1(f)$ on the image domain $P$ can still be improved. In the following steps, we show how this can be achieved by simultaneous processing of adjacent values based on SIMD and in-place processing.
\label{window_decompositon}

\subsection{Utilization of SIMD instructions}
\label{simd_accel}

\begin{figure}
  \centering
    \includegraphics[width=0.45\textwidth]{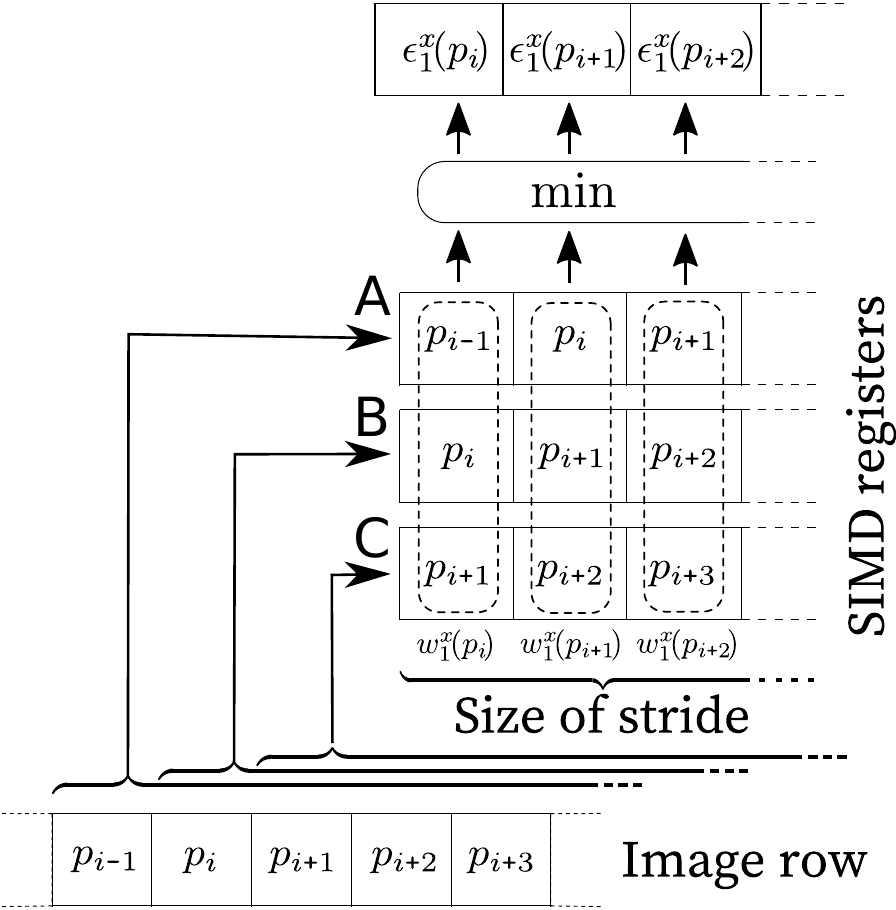}
  \caption{Registers load from positions given by structuring element $w_1^x(p_i)$. Minima are determined from the juxtaposed values in the registers, thereby, assuming stride size 4, giving simultaneous computation $\epsilon_1^x(f)$ from $p_i$ to $p_{i\texttt{+}3}$.}
  \label{fig:simd_load_horizontal}
\end{figure}

The amount of simultaneous computation is determined by the data type and the SIMD register size, which depends on the targeted instruction set architecture. This defines the size of processing stride and, thus, determines how many values are operated on at any given time. The proposed method is described using 256-bit SIMD registers (from the AVX2 instruction set) and 64-bit data type, which gives a stride size of $4$. 
When $[\epsilon_1^x(f)](p_i)$ is applied, values from the range $[p_{i\texttt{-}1}, p_{i\texttt{+}2}]$ are read into the register A at once, while registers B and C retrieve values from the range $[p_{i}, p_{i\texttt{+}3}]$ and  $[p_{i\texttt{+}1}, p_{i\texttt{+}4}]$, respectively. In this way, SIMD registers act as displaced queues of values, allowing for simultaneous computation of $\epsilon_1^x(f)$ from the range $[p_{i}, p_{i\texttt{+}3}]$ by determining the minima of juxtaposed values of registers A, B and C, as shown in Fig. \ref{fig:simd_load_horizontal}. Obviously, $\epsilon_1^y(f)$ is applied in the same manner, switching from row-wise to column-wise offsets. As $4$ comparisons are carried out at once, this improves the computational efficiency significantly, however, the system memory load is increased, as $256$ bits of data are processed in all cases. This is mitigated by in-place processing, as described next.

\subsection{In-place processing}

\begin{figure}
  \centering
    \includegraphics[width=0.45\textwidth]{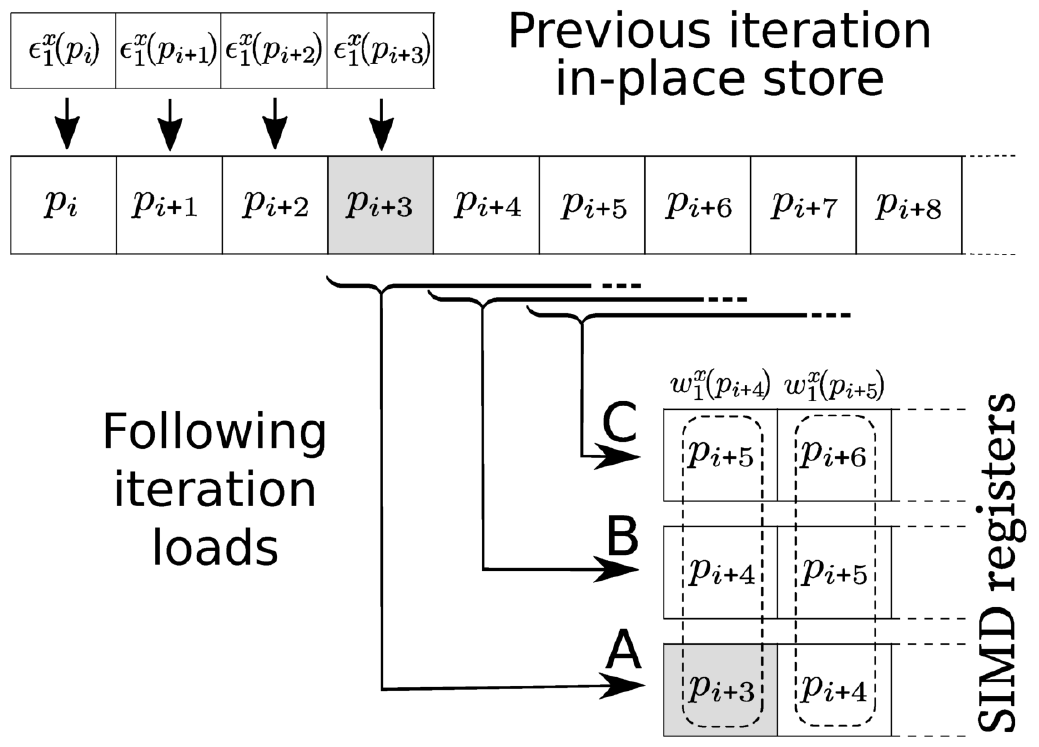}
  \caption{The overlapping value (highlighted in gray) between SIMD strides must be preserved for following evaluation before overriding it with the results the of the previous iteration.}
  \label{fig:simd_overlap}
\end{figure}

The computation of $\epsilon_1^x(f)$ cannot be stored in-place directly, due to the overlapping of structuring elements. Namely, when using $w_1^x$, the last value processed during each iteration is considered during the following iteration, as shown in Fig. \ref{fig:simd_overlap}. In order to preserve the overlapping value, values are read into register A before the computed minimum values are stored in-place. Registers B and C are read afterwards, in order to use only three registers at once. The pseudocode is provided by Algorithm \ref{algo:horizontal_erode}, where function $\text{Load}_L(f, idx)$ loads $L$ pixel values from image $f$ at linear offset $[idx, idx+L-1]$ and, conversely, $\text{Store}_L(f, idx)$ updates $L$ image values from a given register, accordingly.
\begin{algorithm}
\SetAlgoLined
\SetKw{KwBy}{by}
\SetKwFunction{Load}{$\text{Load}_L$}
\SetKwFunction{Store}{$\text{Store}_L$}
\SetKwFunction{lerode}{inplace\_row\_erode}
\DontPrintSemicolon

\SetKwProg{Fn}{Function}{:}{}
  \KwData{image $f$, row offset $y_i$, image width $X$, size of stride $L$}
  \Fn{\lerode{$f$, $y$, $X$, $L$}}{
  \tcp{Preload register A}
  $j \gets X * y$ \;
  $A \gets$ \Load $(f, j-1)$ \;
  \For{$idx \gets j$ \KwTo $j + X - 1$ \KwBy $L$}{
  \tcp{Compute $\epsilon_1^x$}
  $B \gets$ \Load $(f, idx)$\;
  $C \gets$ \Load $(f, idx\texttt{+}1)$\;
  $B \gets \min{A,B,C}$\;
  \tcp{Prepare for next iteration}
  $A \gets$ \Load $(f, idx+L-1)$\;
  \tcp{Store computation in-place}
  $\Store (f, idx) \gets B$
  }
  }
  \caption{In-place sequential computation of $\epsilon_1^x(f)$ for row $y$}
  \label{algo:horizontal_erode}
\end{algorithm}

\begin{figure}
  \centering
    \includegraphics[width=0.35\textwidth]{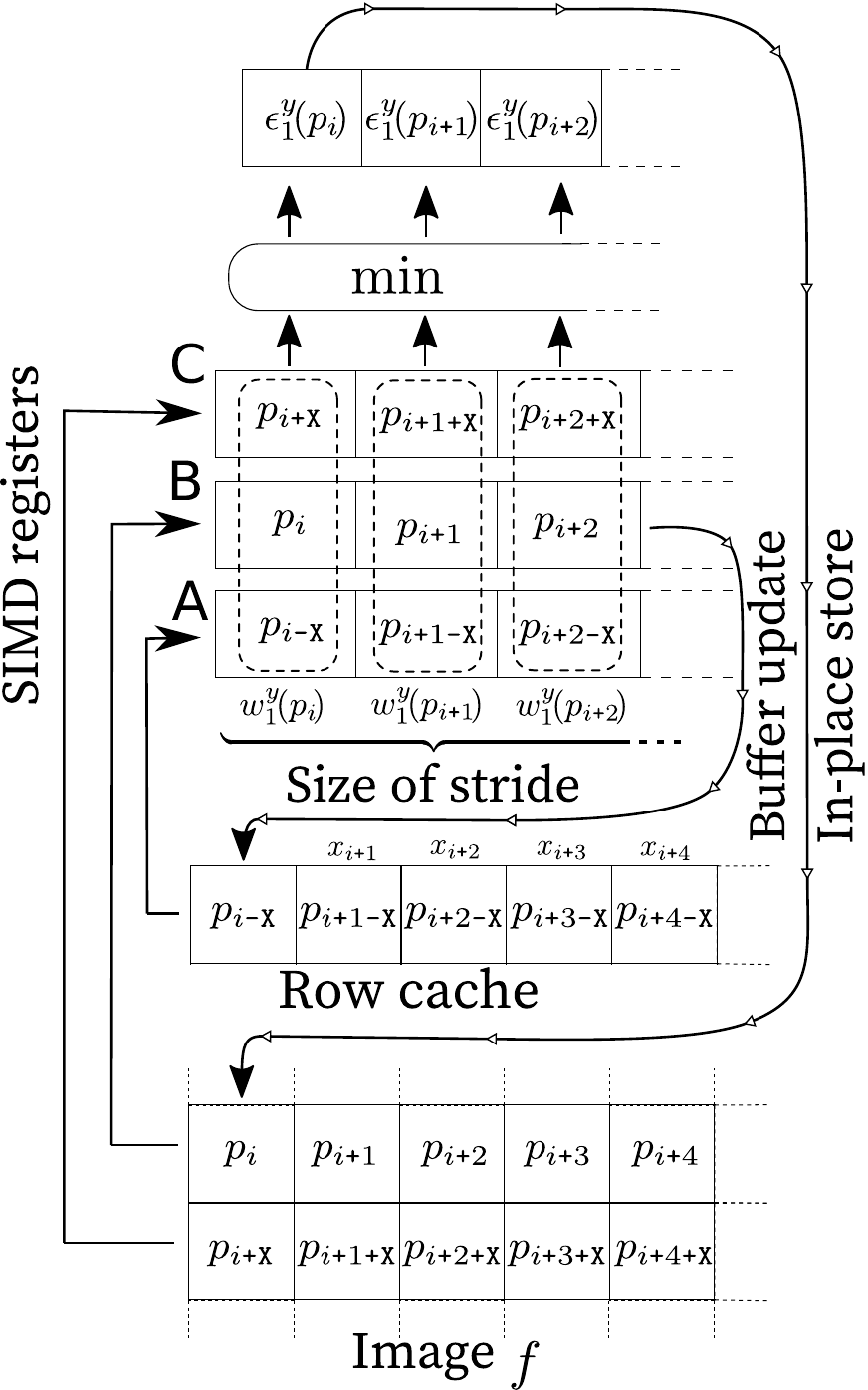}
  \caption{In-place computation of $\epsilon_1^y(f)$ with the use of a single row buffer. Values overwritten by the in-place update are cached in the row buffer to enable computation of the following row.}
  \label{fig:simd_load_vertical_from_horizontal}
\end{figure}

In order to achieve streaming, $\epsilon_1^y(f)$ is resolved in the same sequence as $\epsilon_1^x(f)$. To this end, a single buffer $c$ of size $X$ is created for caching image row values, $c: \mathbb{Z} \to \mathbb{R}$. The buffer is used to preserve row values which were overwritten by previous iterations. Therefore, register A reads values from buffer $c$ at range $[x_{i}, x_{i\texttt{+}3}]$. Conversely, registers B and C read values from image $f$ at ranges $[p_i, p_{i\texttt{+}3}]$ and $[p_{i\texttt{+X}}, p_{i\texttt{+X+3}}]$, respectively. $\epsilon_1^y(f)$ at range $[p_{i}, p_{i\texttt{+}3}]$ is, thus, computed from the minima of juxtaposed register values and stored in-place. For succeeding computations, buffer $c$ is updated using values from register B at their respective range. This is illustrated in Fig. \ref{fig:simd_load_vertical_from_horizontal}, while the pseudocode is provided by Algorithm \ref{algo:column_erosion}. The following subsection combines $\epsilon_1^x(f)$ and $\epsilon_1^y(f)$ to achieve stream processing of $\epsilon_1(f)$. 
\begin{algorithm}
\SetAlgoLined
\SetKw{KwBy}{by}
\SetKwFunction{Load}{$\text{Load}_L$}
\SetKwFunction{Store}{$\text{Store}_L$}
\SetKwFunction{cerode}{inplace\_vertical\_erode}
\DontPrintSemicolon
\SetKwProg{Fn}{Function}{:}{}
  \KwData{image $f$, image width $X$, image height $Y$, size of stride $L$}
  \Fn{\cerode{$f$, $X$, $Y$, $L$}}{
  \tcp{Initialize buffer}
  $c \gets \{\max{\mathbb{R}},\dots,\max{\mathbb{R}}\}$\;
  
  \For{$row\gets0$ \KwTo $Y-1$}{
  \For{$column\gets0$ \KwTo $X-1$ \KwBy $L$}{
  $idx \gets row * X  + column$ \;
  \tcp{Compute $\epsilon_1^y$ in-place}
  $A \gets \Load (c, column)$\;
  $B \gets \Load (f, idx)$\;
  $C \gets \Load (f, idx+X)$\;
  $A \gets \min{A,B,C}$\;
  $\Store (f, idx) \gets A$\;
  \tcp{Update buffer}
  $\Store (c, column) \gets B$\;
  }
  }
  }
  \caption{In-place sequential computation $\epsilon_1^y(f)$ across image domain $P$.}
  \label{algo:column_erosion}
\end{algorithm}

\subsection{Stream processing}
\label{sec:streaming_kernel}
\begin{figure}
  \centering
    \includegraphics[width=0.45\textwidth]{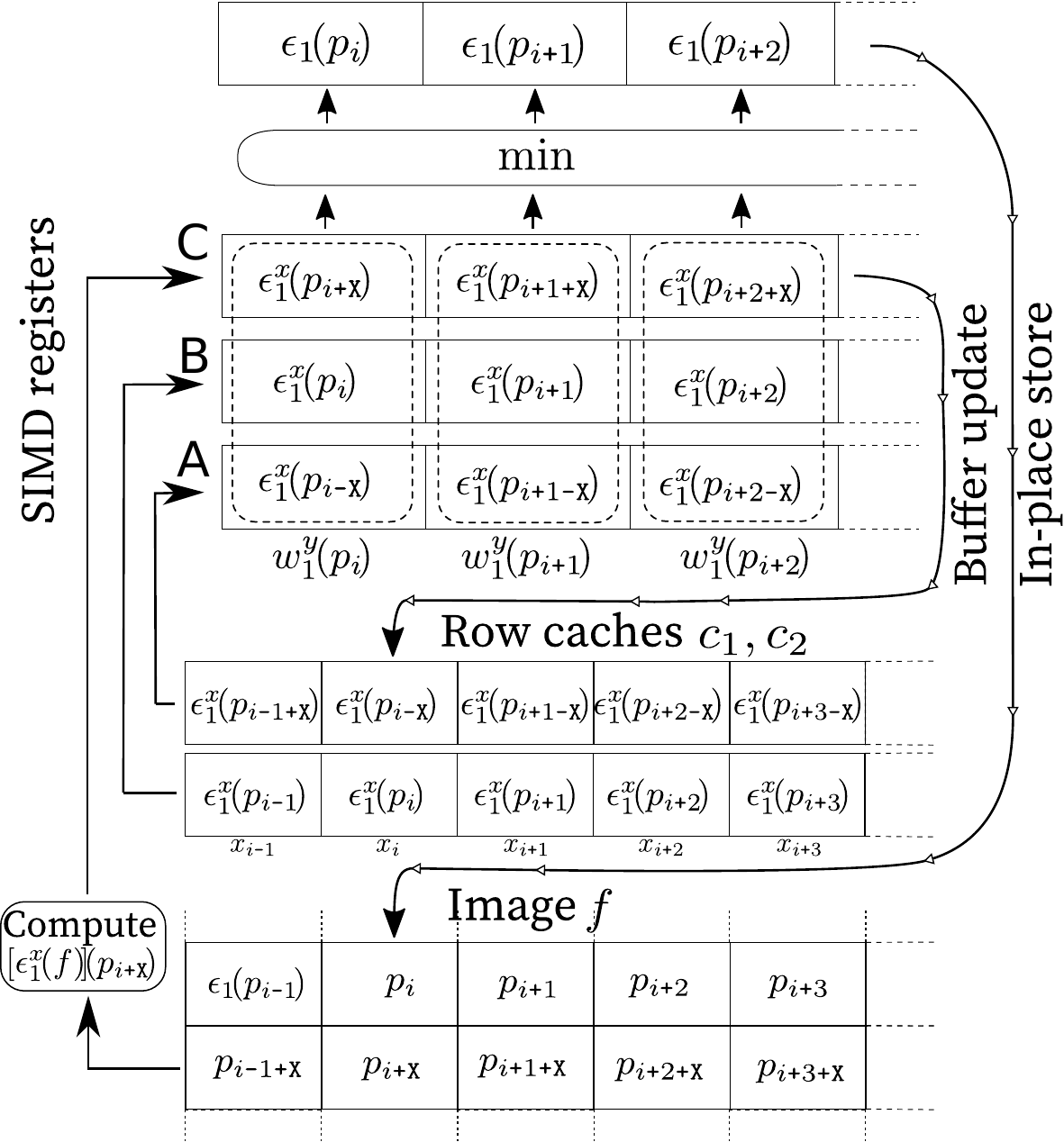}
  \caption{In-place $[\epsilon_1(f)](p_i)$, where two row buffers are used to enable stream processing. Cached values in the buffer are updated for further sequential computations.}
  \label{fig:simd_load_vertical}
\end{figure}

In order to achieve streaming of $\epsilon_1(f)$ while retaining computational efficiency, two buffers are introduced for caching $c_1, c_2: \mathbb{Z} \to \mathbb{R}$ of size $X$. These are used to cache computations of $\epsilon_1^x(f)$ of the previous two rows. Therefore, registers A and B read values at range $[x_{i}, x_{i\texttt{+}3}]$ from $c_1$ and $c_2$, respectively. $\epsilon_1^x(f)$ is computed from the image at range $[p_{i\texttt{+X}}, p_{i\texttt{+X+3}}]$ and stored in register C. It follows that the juxtaposed minima of registers gives $\epsilon_1(f)$ at range $[p_{i}, p_{i\texttt{+}3}]$, which are written in-place. For succeeding computations, cache $c_1$ is updated using register C values at range $[x_{i}, x_{i\texttt{+}3}]$. Representations of buffers $c_1$ and $c_2$ are swapped once an entire image row is processed. The entire procedure is illustrated in Fig. \ref{fig:simd_load_vertical}. Elementary geodesic erosion is implemented trivially by constraining $\epsilon_1(f)$ with corresponding values from mask $m$, as described by Algorithm \ref{algo:geodesic_erosion}. The following subsections describe the use of streamable filters in filter chains and their efficient parallel computation with available multicore resources.
\begin{algorithm}
\SetAlgoLined
\SetKw{KwBy}{by}
\SetKwFunction{Swap}{Swap}
\SetKwFunction{Load}{$\text{Load}_L$}
\SetKwFunction{Store}{$\text{Store}_L$}
\SetKwFunction{geoerode}{geodesic\_erode}
\DontPrintSemicolon

\SetKwProg{Fn}{Function}{:}{}
  \KwData{image $f$, mask image $m$, image width $X$, image height $Y$, size of stride $L$}
  \Fn{\geoerode{$f$, $m$, $X$, $Y$, $L$}}{
  \tcp{Initialize buffers}
  $c_1 \gets \{\max{\mathbb{R}},\dots,\max{\mathbb{R}}\}$\;
  $c_2 \gets \{[\epsilon_1^x(f)](p_0),\dots,[\epsilon_1^x(f)](p_{X-1})\}$\;
  
  \For{$row\gets0$ \KwTo $Y-1$}{
  \For{$column\gets0$ \KwTo $X-1$ \KwBy $L$}{
  $idx \gets row * X  + column$ \;
  \tcp{Compute $\epsilon_1$}
  $A \gets \Load (c_1, column)$\;
  $B \gets \Load (c_2, column)$\;
  $C \gets \Load (\epsilon_1^x(f), idx+X)$\;
  $A \gets \min{A,B,C}$\;
  \tcp{Constrain $\epsilon_1$ with $m$}
  $B \gets \Load(m,idx)$\;
  $\Store (f, idx) \gets \max{A,B}$\;
  \tcp{Update buffer}
  $\Store (c_1, column) \gets C$\;
  	}
  	$\Swap(c_1, c_2)$
  }
  }
  \caption{In-place sequential computation of $\epsilon_{1}^{m}(f)$ across image domain $P$.}
  \label{algo:geodesic_erosion}
\end{algorithm}
\subsection{Efficient parallel processing}
\label{sub:multicoreprocessing}
In order to process a chain of streamable filters efficiently, a processing pipeline is established, which consists of a per worker thread task queue and a thread pool, used to manage worker threads. During the initialization stage, the thread pool is populated by creating $T$ worker threads, where $T$ is defined by the available multicore resources on a given CPU. Each thread is then paired with its own task queue, and it suspends itself when its task queue is emptied. During the run-time, each thread is responsible solely for computing its queued filters. Thus, whenever a filter chain is constructed it is enqueued sequentially into threads' task queues using modulo operation. That is, the filter at index $j$ is enqueued into the $t$-th thread's task queue, where ${t = 1 + (j-1) \;\mathrm{mod}\; T}$. Finally, the thread is resumed if its corresponding task queue was previously empty.

As $T$ filters are computed simultaneously in this way, their synchronization is required in order to mitigate data race conditions in regard to each other. This is achieved by tracking the number of processed rows by a given filter $j$ (i.e. a thread) using the atomic counter $r_j$ associated with it. Note that $r_j$ is incremented after an image row has been processed, making the granularity of synchronization per image row. In order to prevent data races and ensure sequential application of filters, filter $j$ reads image values at offset $idx$ only when filter $j-1$ has processed that row fully and incremented its atomic counter $r_{j-1}$. This constraint synchronizes the entire sequence of filters, where the implementation utilizes a spinlock for synchronization on atomic counters (e.g. lines 6 to 8 in Algorithm \ref{algo:geodesic_erosion_by_reconstruction}), though other synchronization primitives could also be used (e.g. conditional variables).
Finally, acquire-release semantics guarantees that any data which are written before the atomic row counter is modified will be synchronized implicitly when the same row counter is read in another thread \cite{poter2018memory}. 

\subsection{Maximizing data locality}
\begin{figure}[h]
  \centering
    \includegraphics[width=0.48\textwidth]{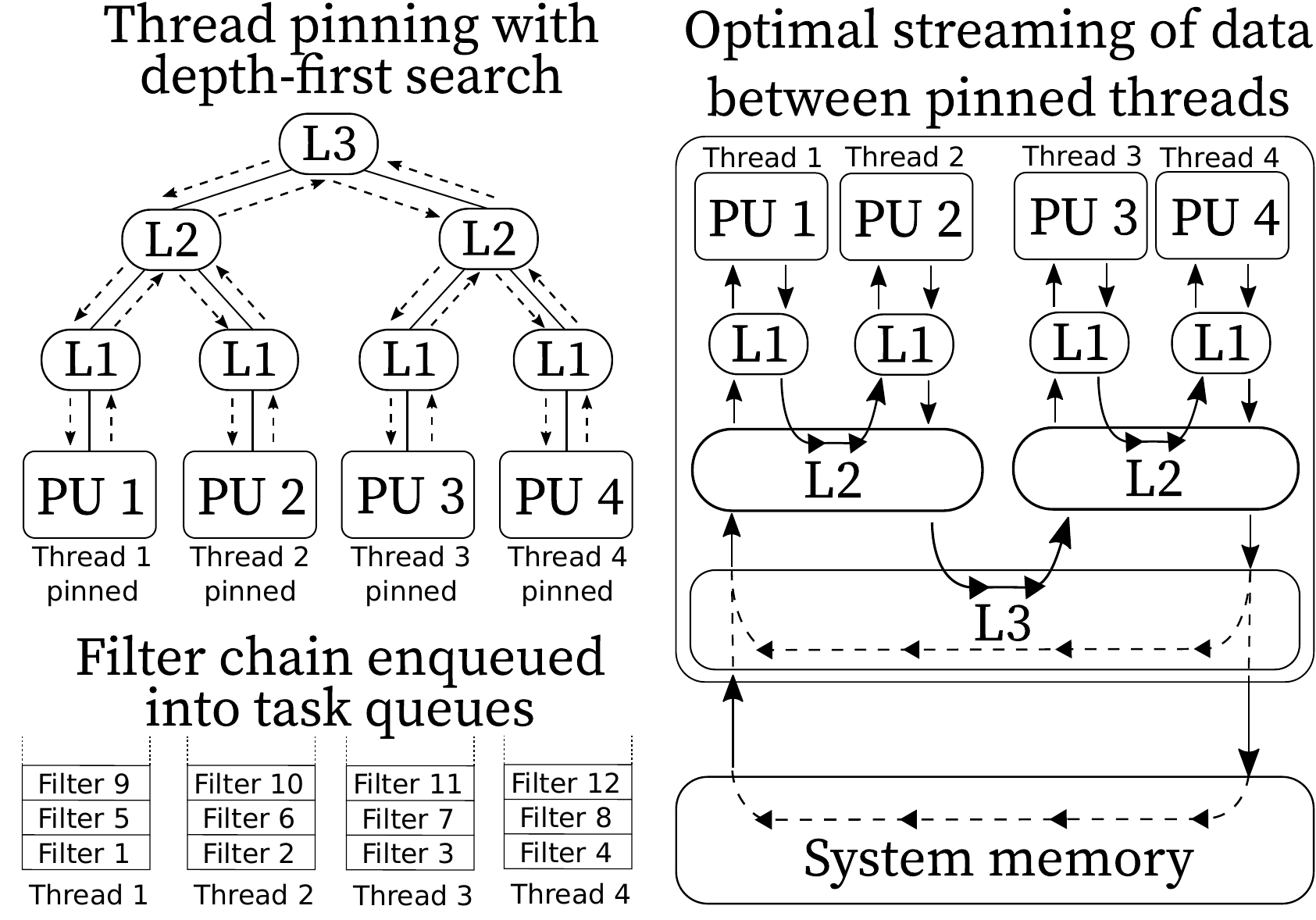}
  \caption{Example of processing a filter chain on a hypothetical quad-core configuration by leveraging thread pinning to maximize data locality via the CPU cache.}
  \label{fig:streaming_ideal}
\end{figure}

The topology of the CPU cache is leveraged in order to make the implicit image data synchronization more efficient. Note that the output of filter $j-1$ is not stored directly to main memory, but is first stored transiently by the CPU cache. As this is the direct input of filter $j$, the threads computing them are kept neighboring in regard to the CPU cache. The optimal thread pinning is determined at the initialization stage, using a depth-first search to traverse the system topology tree, which is obtained with the hwloc library \cite{broquedis2010hwloc}. Threads are pinned sequentially when a PU (Processing Unit) leaf node is reached in the topology tree. This increases the amount of data transfers occurring within a higher level cache, thereby mitigating for the decreased bandwidth and higher latency of lower level caches or system memory \cite{gummaraju2004stream, mazouz2011performance}. An idealized example of using thread pinning is shown in Fig. \ref{fig:streaming_ideal}, where data transfer between threads executing on PU1 and PU2 (PU3 and PU4) occurs via the L2 cache, while the L3 cache is used to transfer data between PU2 and PU3. When L3 is sufficiently large to hold all the image data, it is also used for data transfer between PU4 and PU1, otherwise system memory is used. Additional implementation details of reconstruction filters and quasi-distance transformation are described next. 

\subsection{Implementation details of geodesic operators}
\label{subsection:reconstruction_qdt_impl_details}
The geodesic operators from Eqs. \ref{eq:hmaxima}-\ref{eq:openbyrecon} all feature reconstruction filters. These require a run-time mechanism to detect convergence, as it is not possible to determine, in advance, the length of filter chain required for erosion by reconstruction $\epsilon_{\text{rec}}^{m}$ (or dilation by reconstruction $\delta_{\text{rec}}^{m}$). To this end, reconstruction filters test for convergence, and, until it is achieved, $\epsilon^{m}_{1}$ ($\delta^{m}_{1}$) requeues itself into the front of the thread's task queue, as described by Algorithm \ref{algo:geodesic_erosion_by_reconstruction}. Using a chain length of $T$ is, therefore, sufficient to occupy all available multicore resources and guarantee convergence. Implementation of $\delta_{\text{rec}}^{m}$ is apparent from the given details of $\epsilon_{\text{rec}}^{m}$, while the remaining geodesic operators from Eqs. \ref{eq:hmaxima}-\ref{eq:openbyrecon} are then derived using basic image operations.
\begin{algorithm}
\SetAlgoLined
\SetKw{KwBy}{by}
\SetKwFunction{Swap}{Swap}
\SetKwFunction{Load}{$\text{Load}_L$}
\SetKwFunction{Store}{$\text{Store}_L$}
\SetKwFunction{Requeue}{Requeue}
\SetKwFunction{Wait}{Wait}
\SetKwFunction{geoerode}{stream\_geo\_erode}

\DontPrintSemicolon

\SetKwProg{Fn}{Function}{:}{}
  \KwData{marker image $f$, mask image $m$, image width $X$, image height $Y$, size of stride $L$, sequential filter id $j$, number of threads $T$}
  \Fn{\geoerode{$f$, $m$, $X$, $Y$, $L$, $j$}}{
  \tcp{Initialize caches}
  $c_1 \gets \{\max{\mathbb{R}},\dots,\max{\mathbb{R}}\}$\;
  $c_2 \gets \{[\epsilon_1^x(f)](p_0),\dots,[\epsilon_1^x(f)](p_{\texttt{X-}1})\}$\;
  
  \tcp{Assume convergence}
  $converged \gets $ true\;
  \For{$row\gets0$ \KwTo $Y-1$}{
  \While{$row \geq r_{j\texttt{-}1}$}{
	\Wait  
  }
  
  \For{$column\gets0$ \KwTo $X-1$ \KwBy $L$}{ 
  $idx \gets row * X  + column$ \;
  \tcp{Compute $\epsilon_1$}
  $A \gets \Load(c_1, column)$\;
  $B \gets \Load(c_2, column)$\;
  $C \gets \Load(\epsilon_1^x(f), idx+X)$\;
  $A \gets \min{A,B,C}$\;
  \tcp{Update cache}
  $\Store(c_1,column) \gets C$\;
  \tcp{Constrain $\epsilon_1$ with $m$}
  $B \gets \Load(m, idx)$\;
  $A \gets \max{A,B}$\;
  \tcc{Compare to previous values in $f$}
  $B \gets \Load(f, idx)$\;
  \If{$A \neq B$}{
  	$\Store(f, idx) \gets A$\;
  	\tcp{Reject assumption}
  	$converged \gets$ false\;
  }
  	}
  \tcc{Increment row counter and swap cache rows}
    $r_j \gets r_j + 1$\;
  	\Swap{$c_1$, $c_2$}
  }
  \tcp{Requeue if $f$ was updated}
  \If{$converged \neq$ true}{
  	 \Requeue{\geoerode{f, m, X, Y, L, j+T}}
  }
  }
  \caption{In-place $\epsilon_{1}^{m}$ with convergence detection, suitable for use in reconstruction filters.}
  \label{algo:geodesic_erosion_by_reconstruction}
\end{algorithm}

Quasi-distance transformation (Eq. \ref{eq:quasi_eq1}) requires computation of increasing erosions to obtain the largest residuals and their geodesic distance, as given by Algorithm \ref{algo:qdt_erosion}. Increasing erosions are applied iteratively with $\epsilon_1$, where $d(f)$ and $r(f)$ are updated accordingly. The maximum length of required filter chain is given by the larger image dimension $\max{(X,Y)}$. In order to minimize image writes, an additional conditional statement is used to test for changes in $r(f)$ (and, consequently, $d(f)$), instead of storing the maximum straightforwardly between the previous and the current values (see lines 15 to 23 in Algorithm \ref{algo:qdt_erosion}). Despite these conditional updates decreasing the effectiveness of speculative execution, the results confirmed that the lower bandwidth requirements outweigh this drawback (see Section \ref{subsection:results_geodesic_ops}). Implementation of $d_{L1}(f)$ is apparent from the given details of $\epsilon_{\text{rec}}^{m}$ and Eq. \ref{eq:qdt_eq2}, as it is a specialized case of reconstruction filter.  
\begin{algorithm}
\SetAlgoLined
\SetKw{KwBy}{by}
\SetKwFunction{Swap}{Swap}
\SetKwFunction{Load}{$\text{Load}_L$}
\SetKwFunction{Store}{$\text{Store}_L$}
\SetKwFunction{Enqueue}{Enqueue}
\SetKwFunction{Wait}{Wait}
\SetKwFunction{MaskedStore}{$\text{\small{MaskedStore}}_L$}
\SetKwFunction{FQdt}{erode\_QDT}
\DontPrintSemicolon
  
\SetKwProg{Fn}{Function}{:}{}
  \KwData{input image $f$, delta image $r$, distance image $d$, image width $X$, image height $Y$ size of stride $L$, sequential filter id $j$}
  \Fn{\FQdt{$f$, $r$, $d$, $X$, $Y$, $L$, $j$}}{
  $c_1 \gets \{\max{\mathbb{R}},\dots,\max{\mathbb{R}}\}$\;
  $c_2 \gets \{[\epsilon_1^x(f)](p_0),\dots,[\epsilon_1^x(f)](p_{X-1})\}$\;
  
  \For{$row\gets0$ \KwTo $Y-1$}{
  \While{$row \geq r_{j-1}$}{
	\Wait  
  }
  
  \For{$column\gets0$ \KwTo $X-1$ \KwBy $L$}{   
  $idx \gets row * X  + column$ \;
  \tcp{Compute $\epsilon_1$}
  $A \gets \Load(c_1, column)$\;
  $B \gets \Load(c_2, column)$\;
  $C \gets \Load(\epsilon_1^x(f), idx+X)$\;
  $A \gets \min{A,B,C}$\;           \tcp{Update cache}
  $\Store(c_1, column) \gets C$\;
  \tcp{Compute iterative residuals}
  $B \gets \Load(f, idx)$\; $B \gets B-A$\;
  \tcp{Update image $f$ with $\epsilon_1$}
  $\Store(f, idx) \gets A$\;
  \tcc{Update $r(f)$ and $d(f)$ only where bitmask is true}
  $A \gets \Load(r(f), idx)$\;
  $mask \gets B > A$\; \If{$mask \neq 0$}{
  ${\MaskedStore(r(f),idx,mask)\!\gets\!B}$\;
  ${B \gets \{j,\dots, j\}}$\;
  ${\MaskedStore(d(f),idx,mask)\!\gets\!B}$\;
  }
  	}
    $r_j \gets r_j + 1$\;
  	\Swap{$c_1$, $c_2$}\;
  }
  \caption{Extended $\epsilon_1$ filter which, with use of $r(f)$, computes discontinuous function $d(f)$ of quasi-distance transform.}
  \label{algo:qdt_erosion}
  }
  
\end{algorithm}

\section{Results}
\label{sec:results}
\begin{table*}
\begin{center}
    \caption{Test systems used for performance evaluations}
    \resizebox{\textwidth}{!}{\begin{tabular}{ c c c c c c c c c c c c c }
    \toprule
    \multirow{2}{*}{\shortstack[c]{Test\\system}} & \multicolumn{3}{c}{CPU}  &  \multirow{2}{*}{\shortstack[c]{Num. of \\sockets}} & \multicolumn{3}{c}{Total CPU cache per socket} & &\multicolumn{3}{c}{GPU}\\ \cmidrule{2-4} \cmidrule{6-8} \cmidrule{10-12}
    \multirow{2}{*}{}& Vendor & Model & Frequency & \multirow{2}{*}{} & L1D & L2 & L3 & &Nvidia & CUDA cores & Peak GFLOPS \\  \midrule
    1 & Intel  & Core i5-6600k  & 3.4GHz & 1 & 128KiB & 1MiB & 6MiB && / & / & / \\ 
    2 & AMD & Threadripper 1920X  & 3.5GHz & 1 & 384KiB & 6MiB & 32MiB && GeForce 750Ti & 640 & 1389 \\ 
    3 & Intel & Xeon E5-2683 v4 & 2.10GHz & 2 & 512KiB & 4MiB & 40MiB && Titan X Pascal & 3840 & 10970 \\ \bottomrule
    \end{tabular}}
    \label{table:system_specs}
\end{center}
\end{table*}
\begin{table*}
\begin{center}
    \caption{Cache topology of test systems}
    
    \resizebox{\textwidth}{!}{\begin{tabular}{ ccc c cc c cc c cc }
    \toprule
    \multirow{2}{*}{\shortstack[c]{Test\\system}} & \multirow{2}{*}{\shortstack[c]{PU configuration\\ per socket}} & \multirow{2}{*}{SMT} & \multicolumn{9}{c}{Cache ownership}\\ \cmidrule{5-12}
    \multirow{2}{*}{} & \multirow{2}{*}{} & \multirow{2}{*}{} && L1D & Shared with \#PUs &&  L2 & Shared with \#PUs && L3 & Shared with \#PUs \\ \cmidrule{1-3} \cmidrule{5-6} \cmidrule{8-9} \cmidrule{11-12}
    1 & 4 & / && 32KiB & 1 && 256KiB & 1 && 6MiB & 4 \\ 
    2 & 6+6+6+6 & 2-way && 32KiB & 2 && 512KiB & 2 && 8MiB & 6 \\ 
    3 & 32 & 2-way && 32KiB & 2 && 256KiB & 2 && 40MiB & 32 \\ \bottomrule
    \end{tabular}}
    \label{table:cache_specs}
\end{center}
\end{table*}
\subsection{Test systems}
\label{subsection:test_system}

Three different testing systems were used in order to test method performances on different x86-64 CPU architectures. They all differ in cache topology and the interconnectivity of processing units. However, they all support the Advanced Vector Extensions 2 (AVX2) instruction set, that allows for processing of all data types using 256-bit registers.
The first system was a quad-core CPU with three level cache, which does not support Simultaneous Multi-Threading (SMT). It is used as the baseline for evaluating improvements achieved by the proposed implementation of geodesic operators. Next, a system with Non-Uniform Memory Access (NUMA) architecture was considered. It consisted of four interconnected triple-core processors in a single CPU socket. Since the CPU is fragmented into sub-processors, this provided relevant information with regard to the scaling efficiency. Each sub-processor has its own dedicated 8MiB L3 cache, which is shared among all sub-processors. The last system was also a NUMA architecture, but with two separate monolithic CPUs, each consisting of 16-cores and allowing for SMT. Due to the use of dual sockets, insights were obtained into the scaling efficiency achieved by multiple CPUs. A detailed overview of each system and its cache topology is given in Tables \ref{table:system_specs} and \ref{table:cache_specs}. Performance enhancement techniques, such as frequency boosting at lower thread counts and SMT, were kept enabled on all test systems.

\subsection{Validation protocol}

As running times of geodesic operators depend on the content of the processed image as well as their input parameters, the initial validation focuses on filter chains of fixed length, while actual geodesic operators are examined in Section \ref{subsection:results_geodesic_ops}. All tests were conducted on a single Male image, except for the latter case, where Airport and Airplane images were considered additionally. All three images were retrieved from \cite{sipi2018}. Unless specified otherwise, image dimensions were kept at $1024 \times 1024$ pixels in all cases. Note, however, that the stride size, together with the internal filter buffers and the memory required to store an image, all depend on the pixel data type. Thus, \texttt{unsigned char}, \texttt{unsigned short}, \texttt{float} and \texttt{double} data types were used in order to test how increasing the number of bytes used for pixels' representation (i.e. one, two, four and eight bytes) influences the computational times. In all cases, the improved utilization of SIMD instructions was achieved by memory-alignment of the image rows. The GCC compiler (version 8.2.1) was used for compilation with the following flags \texttt{-O3 -flto -mavx2}. All evaluations were validated in order to ensure avoidance of data races using ThreadSanitizer with both GCC and Clang (version 7.0.0) compilers. Due to the significant run-time overhead, thread safety validation was performed as an independent test.

In compliance with the given conditions, the evaluations focused on the following three aspects:
\begin{itemize}
\item Evaluation of the overall method's performances was conducted by processing the simplest filter ($\epsilon_1$) using different possible configurations. These include an increasing number of threads, an increasing chain length, and the use of data types of increasing sizes.
\item The method's performance dependency on the image dimensions was performed separately in regard to its width and height. The reason for this is the fact that increasing image width directly increases the size of internal buffers and, consequently, the cache pressure, while increasing height, increases the amount of inter-filter synchronization. The smallest tested image dimension was set at 128, as smaller image widths do not utilize 256-bit SIMD with \texttt{char} data type. In addition, smaller heights reduced the level of parallel processing on test system 3, as the proposed method requires at least 128 image rows to occupy all 64 threads fully.
\item The performance of basic geodesic operators was assessed with all PUs to demonstrate the method's scaling.
\end{itemize}

In the following validation, only those results obtained by the third (fastest) testing system are described in detail, while other test systems are referred to for comparison.
\begin {figure*}[!hbtp]
\centering
\includegraphics{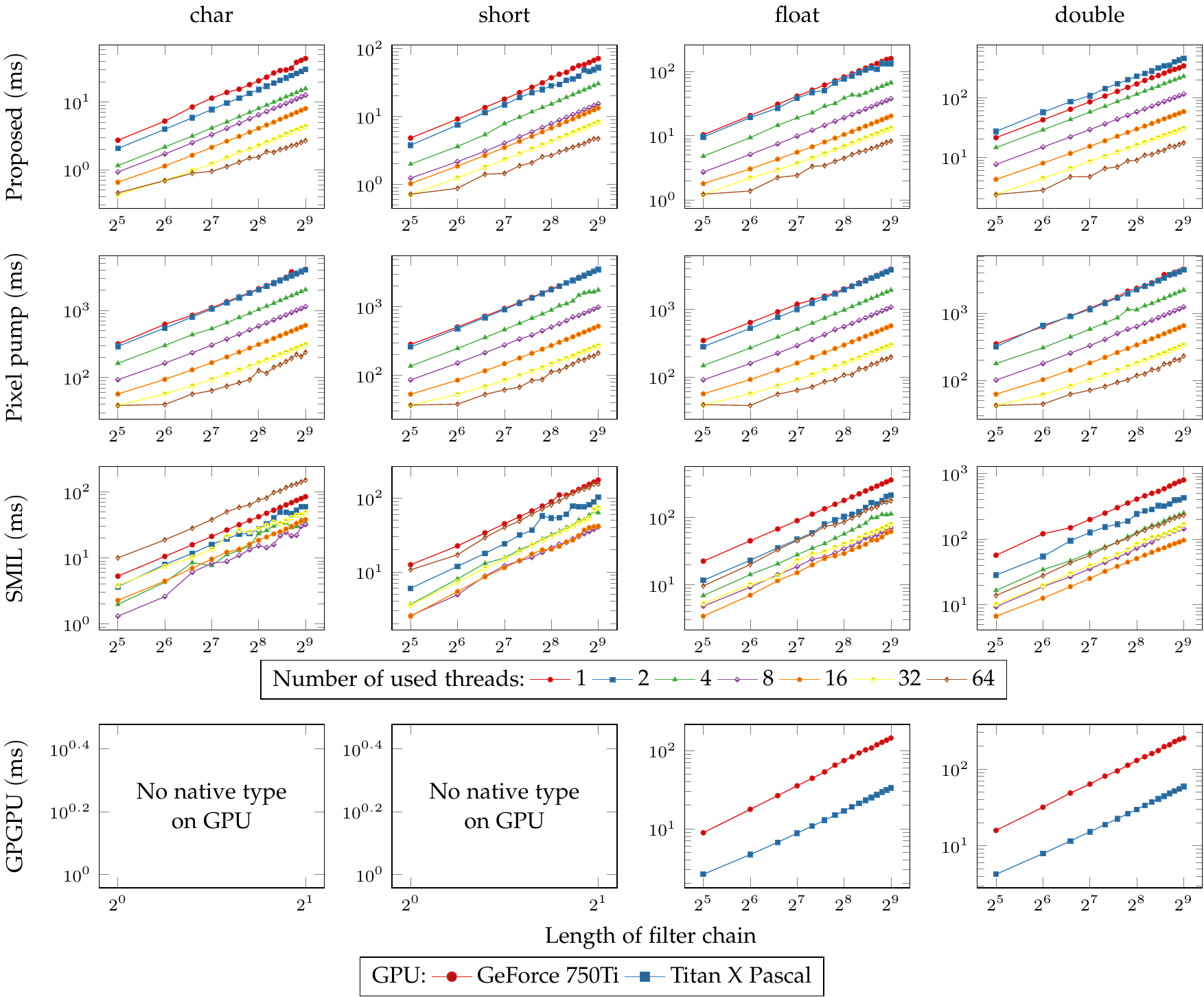}
\caption{Running times of filter chains of varying lengths on Test system 3, where the filter chains are composed of $\epsilon_1$ filters.}
\label{bench:erosion_chain}
\end{figure*}
\begin {figure*}[!hbtp]
\centering
\includegraphics{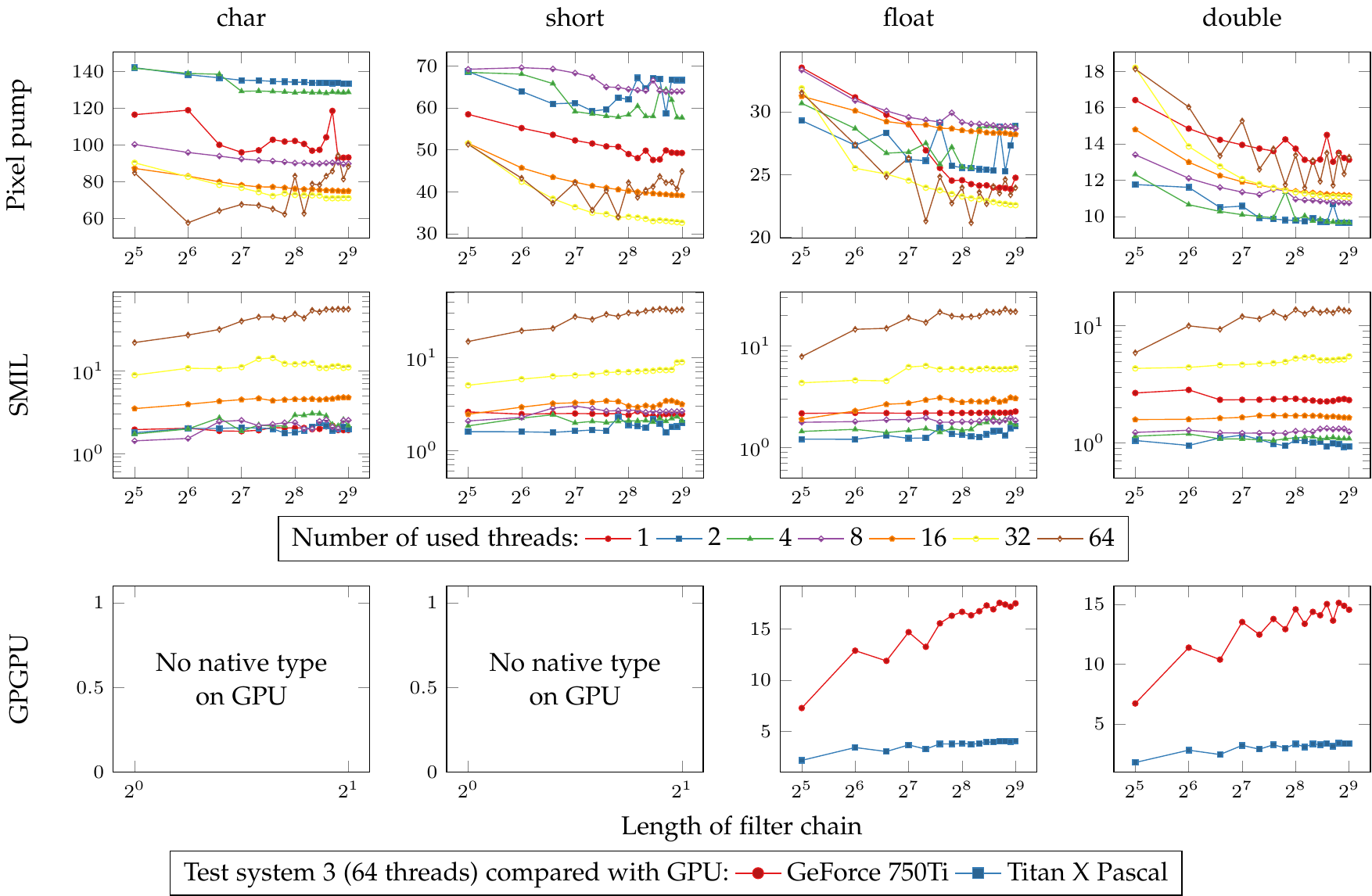}
\caption{Run-time speedup factors achieved by the proposed method on Test system 3 in comparison to pixel pump, SMIL, and GPGPU implementation of filter chains.}
\label{bench:speedups}
\end{figure*}
\subsection{Assessment of time complexity}
\begin{table}
\begin{center}
\setlength\tabcolsep{5pt}
    \caption{Overview of evaluated methods}
    \begin{tabular}{ c c c c r}
    \toprule
    \multicolumn{1}{c}{Method} & \multirow{2}{*}{\shortstack[c]{Comparisons\\per pixel}}  & \multirow{2}{*}{\shortstack[c]{Pipeline\\processing}} & \multirow{2}{*}{\shortstack[c]{Threads\\per filter}} & \multirow{2}{*}{\shortstack[c]{Required\\memory}}\\
    \\
    \midrule 
    Proposed 	& 4  & $T$ filters 	 & 1  & $2X \times T$	\\
    Pixel pump 	& $\BigO{1}$ \cite{dokladal2011computationally} & $T$ filters   & 1 & $(3X\texttt{+}3)\times T$ \\
    SMIL 		& 4  & 1 filter 	& $T$ & $XY$\\
    GPGPU 		& 4  & 1 filter    & $T$ & $XY$\\
    \bottomrule \\
    \multicolumn{5}{l}{\small $X, Y$ - Image width and height, $T$ - Number of PUs}
    \end{tabular}
    \label{table:algo_compare}
\end{center}
\end{table}
In order to assess the overall method's performance, the proposed implementation was compared to the pixel pump streaming algorithm (implemented from pseudo-code provided by \cite{dokladal2011computationally}), the Simple Morphological Image Library (SMIL) \cite{faessel2014smil}, and General-Purpose Graphics Processing Units (GPGPUs) (implemented decomposed structuring element $w_1$ using \cite{podlozhnyuk2007image} as a reference). A comparison of working memory and complexity is shown in Table \ref{table:algo_compare}.

As shown in Fig. \ref{bench:erosion_chain}, the proposed method outperformed the current state-of-the-art streaming algorithm significantly when processing a single filter chain. However, the achieved improvements depend on the data type and the number of threads. As shown in Fig. \ref{bench:speedups}, the largest improvements were achieved when using \texttt{char}, where the speedup factors ranged from $60$ to $130$, while the speedup factors between $9$ and $18$ were achieved in the case of \texttt{double}. This exceeded the maximum speedup from SIMD in both cases, which allowed for simultaneous processing of 	$32$ and $4$ pixels, respectively. The results, thus, demonstrate the method's improved overall use of the CPU facilities, due to the minimization of branching and, consequently, better utilization of the CPU's speculative executions. On the other hand, the method's scaling is sublinear in regard to the number of threads, such as, for example, scaling from $1$ to $64$ threads per filter chain resulted in speedup factors between $16$ to $20$, depending on size of data types. The processing speed of a single filter chain was, therefore, limited by the inter-thread synchronization latency and not inter-thread bandwidth. Speedups achieved when increasing threads per filter chain using the pixel pump algorithm matched that of the proposed method, verifying this fact. 

In single thread evaluations, the proposed method outperformed the SMIL library by speedup factors between $2$ and $4$. A more significant speedup was observed at higher thread counts, where the performances of the SMIL library worsened. As shown in Fig. \ref{bench:speedups}, this is most obvious in the case of $64$ threads, where speedup factors between $10$ and $60$ were achieved. On the other hand, the SMIL library outperformed the proposed method when e.g. using $2$ threads with \texttt{double} data type. This is due to different thread pinning strategies in the presence of SMT, where the proposed method will pin the two threads to the same core, while SMIL uses OpenMP, which will pin threads to separate cores when possible. Without SMT, threads were always pinned to separate cores, where the proposed method also outperformed the SMIL library at low thread counts (see Evaluation 1 of Test system 1 in Table \ref{table:handpicked}).

In all cases, the proposed method also outperformed GPGPU processing of filter chains, with the exception of Test system 1, where the Titan X Pascal GPU displayed better performance. Note, however, as char (1 byte) and short (2 bytes) are not native data types on GPGPU, their evaluation was not examined.

Finally, the comparison of the the achieved pixel throughputs between the proposed and the pixel pump streaming approach is shown in Table \ref{table:handpicked}. Here, it is obvious that the throughput decreased on systems with larger amounts of PUs, as their processing pipelines were longer, causing higher latency per image iteration. The use of SIMD is also apparent, as the throughput decreased, with larger data types. Conversely, the throughput for the pixel pump method remained consistent, due to the scalar processing.

\subsection{Performance dependency on the image dimensions}
\setlength{\abovecaptionskip}{0pt} 
\begin {figure*}[!hbtp]
\centering
\includegraphics{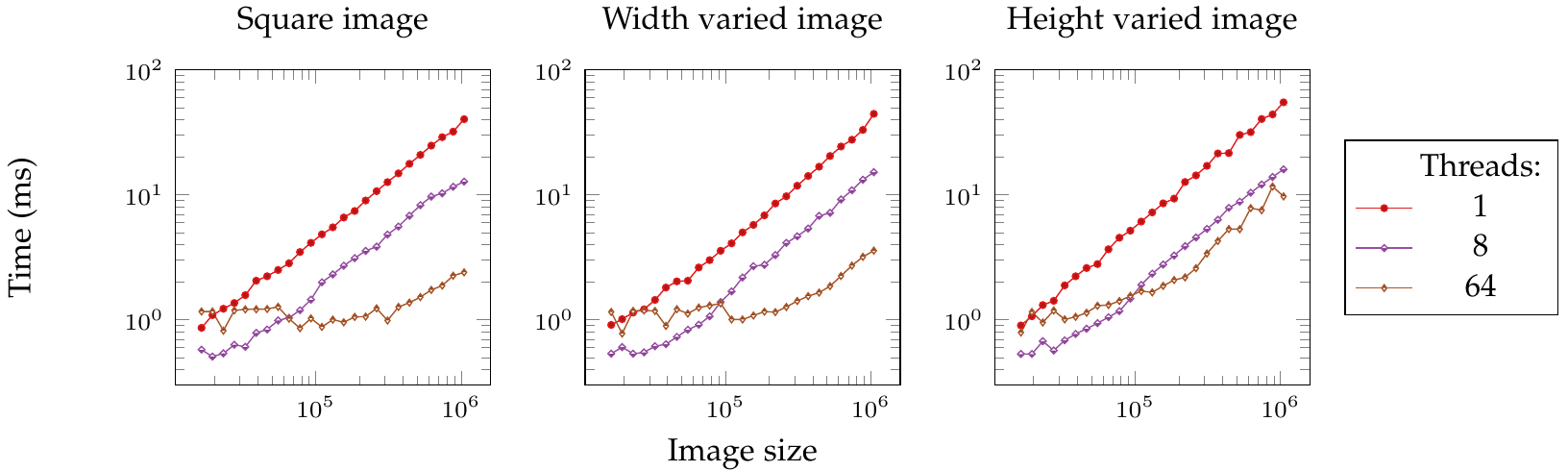}
\caption{Evaluation of the proposed method's performance dependency on image dimensions on test system 3 using a chain of 512 $\epsilon_1$ filters and \texttt{char} data type.}
\label{bench:width}
\end{figure*}
\setlength{\abovecaptionskip}{3pt} 
\begin{table*}[!t]
\begin{center}
\caption{Best running time results of the assessed methods with a 512 $\epsilon_1$ filter chain across tested data types, and effective throughput of streamable methods. The number of threads (CUDA cores for GPGPUs) which gave the result is given in brackets per evaluation.}
\resizebox{\textwidth}{!}{\setlength\tabcolsep{1.5pt}
\begin{tabular}{@{}cc rrrr rrrr rrrr@{}}
\toprule
\multicolumn{2}{l}{}   & \multicolumn{4}{c}{Test system 1}  & \multicolumn{4}{c}{Test system 2} & \multicolumn{4}{c}{Test system 3}                                                                             \\ \cmidrule(lr){3-6} \cmidrule(lr){7-10} \cmidrule(l){11-14}
   & Method & \multicolumn{1}{c}{char} & \multicolumn{1}{c}{short} & \multicolumn{1}{c}{float} & \multicolumn{1}{c}{double} & \multicolumn{1}{c}{char} & \multicolumn{1}{c}{short} & \multicolumn{1}{c}{float} & \multicolumn{1}{c}{double} & \multicolumn{1}{c}{char} & \multicolumn{1}{c}{short} & \multicolumn{1}{c}{float} & \multicolumn{1}{c}{double} \\\cmidrule(r){1-2} \cmidrule(lr){3-6} \cmidrule(lr){7-10} \cmidrule(l){11-14}
\multirow{4}{*}{\begin{tabular}[c]{@{}c@{}}Evaluation 1\\(ms)\end{tabular}}
& Proposed             & 12.9 (4)                 & 17.1 (4)                  & 39.3 (4)                  & 101.4 (4)                  
					   & 9.7 (24)                 & 8.5 (24)                  & 13.8 (24)                 & 28.3 (24)                  
					   & 2.6 (64)                 & 4.7 (64)                  & 8.2 (64)                  & 17.3 (64)                  \\ 
& SMIL                 & 16.1 (4)                 & 29.3 (4)                  & 62.9 (4)                  & 497.2 (4)
                       & 27.7 (6)                 & 35.8 (12)                 & 60.7 (12)                 & 105.3 (12)
                       & 31.9 (8)                 & 40.6 (8)                  & 61.6 (16)                 & 96.0 (16)                 \\ 
& Pixel pump           & 825.0 (4)                & 647.3 (4)                 & 673.0 (4)                 & 797.9 (4)
                       & 269.9 (24)               & 232.1 (24)                & 235.4 (24)                & 271.4 (24)
                       & 237.4 (64)               & 211.0 (64)                & 197.0 (64)                & 231.0 (64)                 \\
& GPU 	               & /		                   & /		                   & /                 		   & /
                       & /			               & /			               & 144.0 (640)               & 253.1 (640)
                       & /                        & /                         & 33.3 (3840)               & 58.9 (3840)                 \\
                       
                       \cmidrule(r){1-2} \cmidrule(lr){3-6} \cmidrule(lr){7-10} \cmidrule(l){11-14}
\multirow{2}{*}{\shortstack[c]{Stream pipeline \\throughput (MPx/s)}} 
&	Proposed	&	9922.48	&	7485.38	&	3257.00	&	1262.33	
		&	2199.31	&	2509.80	&	1545.89	&	753.83	
		&	3076.92	&	1702.13	&	975.61	&	462.43	\\
&	Pixel pump	&	155.15	&	197.74	&	190.19	&	160.42	
		&	79.04	&	91.91	&	90.63	&	78.60	
		&	33.70	&	37.91	&	40.61	&	34.63	\\
                       \cmidrule(r){1-2} \cmidrule(lr){3-6} \cmidrule(lr){7-10} \cmidrule(l){11-14}
\multirow{3}{*}{\shortstack[c]{Evaluation 2\\(ms)}} 
& Square image         & 11.5 (4)                 & 17.1 (4)                  & 38.1 (4)                  & 102.2 (4)
                       & 9.8 (24)                 & 8.5 (24)                  & 13.7 (24)                 & 30.6 (24)
                       & 2.3 (64)                 & 4.8 (64)                  & 8.1 (64)                  & 17.5 (64)                  \\
& Width varied         & 12.9 (4)                 & 30.4 (4)                  & 62.9 (4)                  & 132.2 (4)
                       & 4.4 (24)                 & 8.3 (24)                  & 16.6 (24)                 & 33.7 (24)
                       & 3.5 (64)                 & 7.3 (64)                  & 17.1 (64)                 & 59.2 (32)                  \\
& Height varied        & 13.1 (4)                 & 18.3 (4)                  & 37.0 (4)                  & 107.2 (4)
                       & 17.3 (6)                 & 27.8 (6)                  & 50.0 (6)                  & 45.2 (24)
                       & 4.9 (32)                 & 6.6 (64)                  & 8.3 (64)                  & 14.5 (64)                  \\ \bottomrule
\end{tabular}}
\label{table:handpicked}
\end{center}
\end{table*} Performance dependency on the image dimensions was examined at increasing image sizes, while applying a fixed chain of 512 $\epsilon_1$ filters. As increasing image width increases the buffer size of filters directly, while increasing image height increases the total amount of inter-filter synchronizations, a square image was used as a baseline. Width varied images had their height fixed to $Y=128$, while increasing width $X$ from $128$ to $8192$. Conversely, height varied images had fixed width $X=128$ and increasing height $Y$ from the same range, i.e. $[128, 8192]$. When using \texttt{char} data type, as shown in Fig. \ref{bench:width}, increasing image width did not affect the method's performances significantly, regardless of the numbers of used threads. However, a decrease in the method's performances was noted when increasing the image height. This was because image rows were limited to $128$ pixels, and were, thus, processed promptly. As a result, the majority of processing time was spent on inter-thread synchronization. Using larger data types confirmed this observation, as shown by Evaluation 2 in Table \ref{table:handpicked}. When the data type sizes were increased, the amount of pixels processed simultaneously decreased, due to the smaller strides. This resulted in longer processing times for image rows, and reduced the bottleneck from inter-thread synchronization. From this, it is obvious that image height does not have a significant performance impact when image width is notably larger than the used stride, which is typical in practice. On the other hand, performances worsened by increasing the image width when using larger data types. As each filter's buffer increased in size, it no longer fitted the upper levels of the CPU cache (i.e. L1 and L2) and was, accordingly, moved into the lower levels (i.e. L3). This effect was insignificant in test system 2, due to its larger L2 cache and fragmented L3 cache.

\subsection{Performance assessment of geodesic operators}
\label{subsection:results_geodesic_ops}
\begin{table*}[!ht]
\centering
\caption{Best running time results of basic geodesic operators using \texttt{char} data type (\texttt{float} for GPGPU) and image Male, results are in milliseconds and, where applicable, the number of threads used is given in brackets.}
\resizebox{\linewidth}{!}{
\begin{tabular}{@{}cc rrr  rrrr  rrrr  r@{}}
\toprule
&  & \multicolumn{3}{c}{Test system 1} & \multicolumn{4}{c}{Test system 2} & \multicolumn{4}{c}{Test system 3} & 
	\\\cmidrule(lr){3-5} \cmidrule(lr){6-9} \cmidrule(lr){10-13}
	Operator & \shortstack[c]{Average\\ chain length} & \multicolumn{1}{c}{Proposed} & \multicolumn{1}{c}{SMIL} & \multicolumn{1}{c}{\shortstack[c]{Pixel\\pump}} 
													& \multicolumn{1}{c}{Proposed} & \multicolumn{1}{c}{SMIL} & \multicolumn{1}{c}{\shortstack[c]{Pixel\\pump}} & \multicolumn{1}{c}{\shortstack[c]{GPU}} 
													& \multicolumn{1}{c}{Proposed} & \multicolumn{1}{c}{SMIL} & \multicolumn{1}{c}{\shortstack[c]{Pixel\\pump}} & \multicolumn{1}{c}{\shortstack[c]{GPU}} 
													& \multicolumn{1}{c}{\shortstack[c]{{\scriptsize MCPU}\\\cite{bartovsky2015morphological}}}
	\\\cmidrule(r){1-2} \cmidrule(lr){3-5} \cmidrule(lr){6-9} \cmidrule(lr){10-13} \cmidrule(l){14-14}

HMAX 									& 667 	& 13.3 (4) &	101.9* (4) 	& 1934 (4) & 14.6 (24) &89.2* (6) 	&523.1 (24) & 243.2	& 7.9 (64) &165.5* (8) &483.1 (64) & 55.3 & /\\
DOME 									& 668 	& 14.1 (4) &	102.1* (4) 	& 1934 (4) & 17.9 (24) &89.3* (6) 	&526.7 (24) & 244.9	& 9.3 (64) &164.3* (8) &482.8 (64) & 55.4 & /\\
HFILL 									& 853 	& 17.4 (4) &	106.5* (4) 	& 2906 (4) & 19.1 (24) &90.3* (6) 	&745.5 (24) & 288.7	& 7.4 (64) &169.2* (8) &573.5 (64) & 66.1 & /\\
RAOBJ 									& 1649 	& 31.8 (4) &	104.8* (4) 	& 5655 (4) & 36.1 (24) &86.9* (6) 	&1418 (24)  & 554.4	& 15.4 (64)&168.3* (8) &1016 (64) & 129.6 & /\\
$\gamma_{rec}$	 						& 1353 	& 21.3 (4) &	105.0* (4) 	& 2481 (4) & 28.4 (24) &89.6* (6) 	&716.7 (24) & 463.4	& 16.2 (64)&166.0* (8) &690.0 (64) & 104.6 & /\\
QDT 									& 1140 	& 21.6 (4)& 	/ 			& 1717 (4) & 29.3 (24) &/ 			&615.6 (24) & 652.1 & 7.4  (64)&/ 			&392.3 (64) & 141.3 & /\\
\midrule
$\gamma_{rec}^{75}$						& 1693  & 34.7 (4) &   106.7* (4)  & 4800 (4) & 38.8 (24) &94.0* (6) 	&1488 (24)  & 565.0	& 17.2 (64)&163.6* (8) &772.3 (64) & 129.8 & 544\\
$PS_{[0,11]}$ 							& 77    & 12.9 (4) &   16.1  (4) 	& 194.9 (4) & 4.9 (24)  &16.0 (6) 	&84.5 (24)  & 21.3	& 2.8 (64) &20.6 (8) 	&53.5 (64) & 5.4 & 62.3\\
$ASF_{11}$ 								& 264 	& 4.6 (4)  &   9.9    (4) 	& 465.9 (4) & 5.0 (24)  &14.8 (6) 	&171.4 (24) & 76.6  & 1.6 (64) &18.7 (8) 	&114.6 (64) & 17.5 & 64.2\\
\bottomrule
\\ 
\multicolumn{7}{l}{*The reconstruction algorithm uses a single-threaded hierarchy queue}
\end{tabular}
}
\label{table:geodesic_ops}
\end{table*}
\begin {figure*}[!hbtp]
\centering
\includegraphics{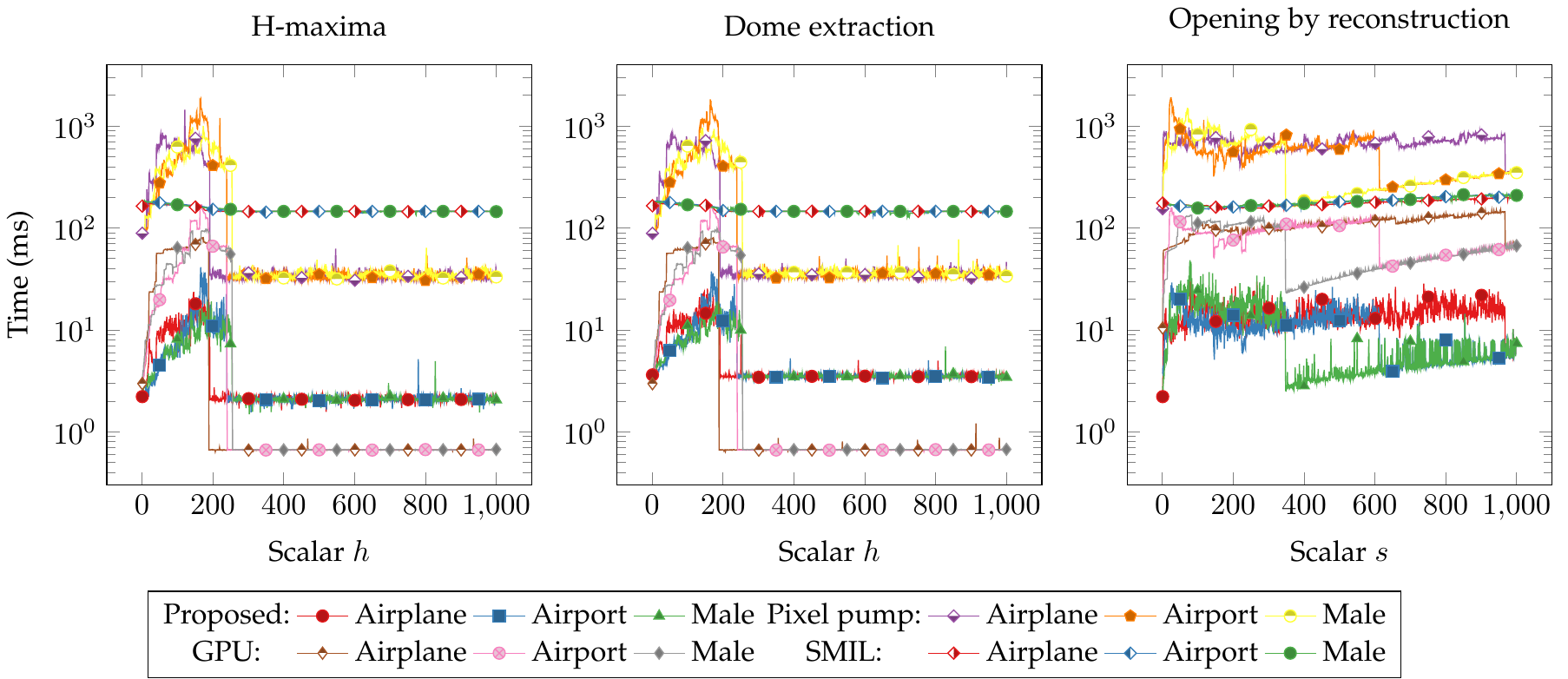}
\caption{Running time results from Test system 3 of geodesic operators with increasing input parameters on images Airplane, Airport and Male \cite{sipi2018}.}
\label{bench:geodesic_comparison}
\end{figure*}
This section provides the assessment of the method's overall performance in practice, when implementing common geodesic operators as defined in Section \ref{sec:theoretical_background}. In those cases where input parameters are required for their computation, the proposed method's sensitivity was assessed by applying it at an increasing scale (see Fig. \ref{bench:geodesic_comparison}), while the obtained average execution times are shown in Table \ref{table:geodesic_ops}. The assessment also includes the results obtained by MCPU based implementation of geodesic operators in order to provide a comparison with a dedicated FPGA architecture. All tests were conducted using \texttt{char} data type, and illustrative outputs of operators are shown in Fig. \ref{fig:basic_geo_ops}.

As expected, the proposed method outperformed the pixel-pump algorithm in all the cases, achieving an average speedup factor of $76$. Still, it is important not to over-stress this fact, as the pixel pump algorithm is dedicated to specialized hardware. Nevertheless, both methods perform notably differently when considering different operators, as the reconstruction process required different lengths of filter chain. In all the cases, however, severe initial filtering achieved when using large input parameters resulted in image degradation, where no reconstruction was possible. 

On the other hand, the SMIL library managed to exhibit near constant performance, regardless of the input parameters, using single-threaded hierarchical queues for reconstruction. When run on Test system 2, the SMIL library outperformed the GPGPU (i.e. GeForce 750Ti) and demonstrated, to a degree, similar results than the GPGPU of Test system 3 (i.e. Titan X Pascal). Still, as shown in Table \ref{table:geodesic_ops}, the proposed method managed to outperform it on all test systems. 

Finally, as shown in Table 5, the proposed method also exceeded the results of MCPU implementation of geodesic operators on all test cases. This can be attributed to the higher frequency of CPUs (up to $3.5$GHz) compared to that of MCPU ($125$MHz), as both required multiple image iterations to process filter chains for the compared geodesic operators fully. This is due to their short pipeline length where, for example, Test system 1 only had 4 PUs and, thus, required $66$ image iterations to apply $ASF_{11}$ fully.

\section{Conclusion}
Following from the results (see Table \ref{table:geodesic_ops}), the proposed method provides real-time processing (30 frames per second) of filter chains composed of over 1500 filters. This was achieved by introducing parallel and SIMD processing within the general data streaming paradigm, where multiple filters were processed in parallel in order to improve data locality. In contrast to the current state-of-the-art implementations of streaming morphological filters (e.g. the pixel pump algorithm) that rely on queue data structures in order to achieve insensitivity to the filter sizes, the proposed method applies filter decomposition for the computation of longer filter chains by using only $3 \times 3$ windows. The SMIL library also utilized SIMD processing, but, for parallel processing, relied on OpenMP, which scaled badly on CPUs with a large amount of PUs. Additionally, the proposed method outperformed even high-end GPUs (see Fig. \ref{bench:speedups} and Table \ref{table:handpicked}) when processing filter chains.

The proposed approach also proved to be more efficient than the comparable streaming method when considering computation of morphological erosions and dilations with window sizes up to $183\times 183$ in the case of using \texttt{char}, and $27\times27$ when using \texttt{double}. However, as the pixel pump method that achieves insensitivity to the filter sizes also supports streaming, integration of both methods into a filter-chain processing framework would allow for optimal performances in all cases. This constitutes our future work. 

\ifCLASSOPTIONcompsoc
\section*{Acknowledgments}
\else
\section*{Acknowledgment}
\fi

The authors acknowledge the financial support from the Slovenian Research Agency (Research Core Funding No. P2-0041 and Project No. J2-8176).

\ifCLASSOPTIONcaptionsoff
  \newpage
\fi

\bibliographystyle{IEEEtran}
\bibliography{IEEEabrv,cites}

  \newpage
\begin{IEEEbiography}[{\includegraphics[width=1in,height=1.25in,clip,keepaspectratio]{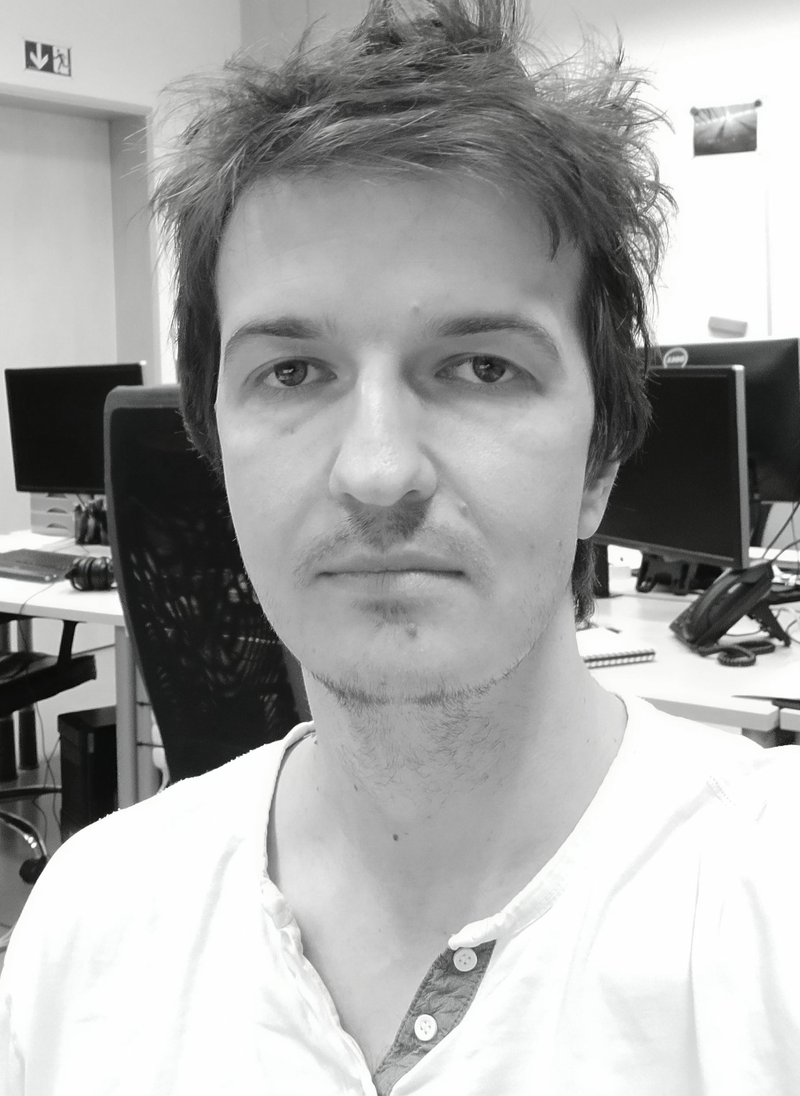}}]{Danijel \v{Z}laus}
is a researcher working towards his PhD degree at the Faculty of Electrical Engineering and Computer Science, University of Maribor. His research interests are Mathematical Morphology, Image Processing and High-Performance Computing.
\end{IEEEbiography}

\begin{IEEEbiography}[{\includegraphics[width=1in,height=1.25in,clip,keepaspectratio]{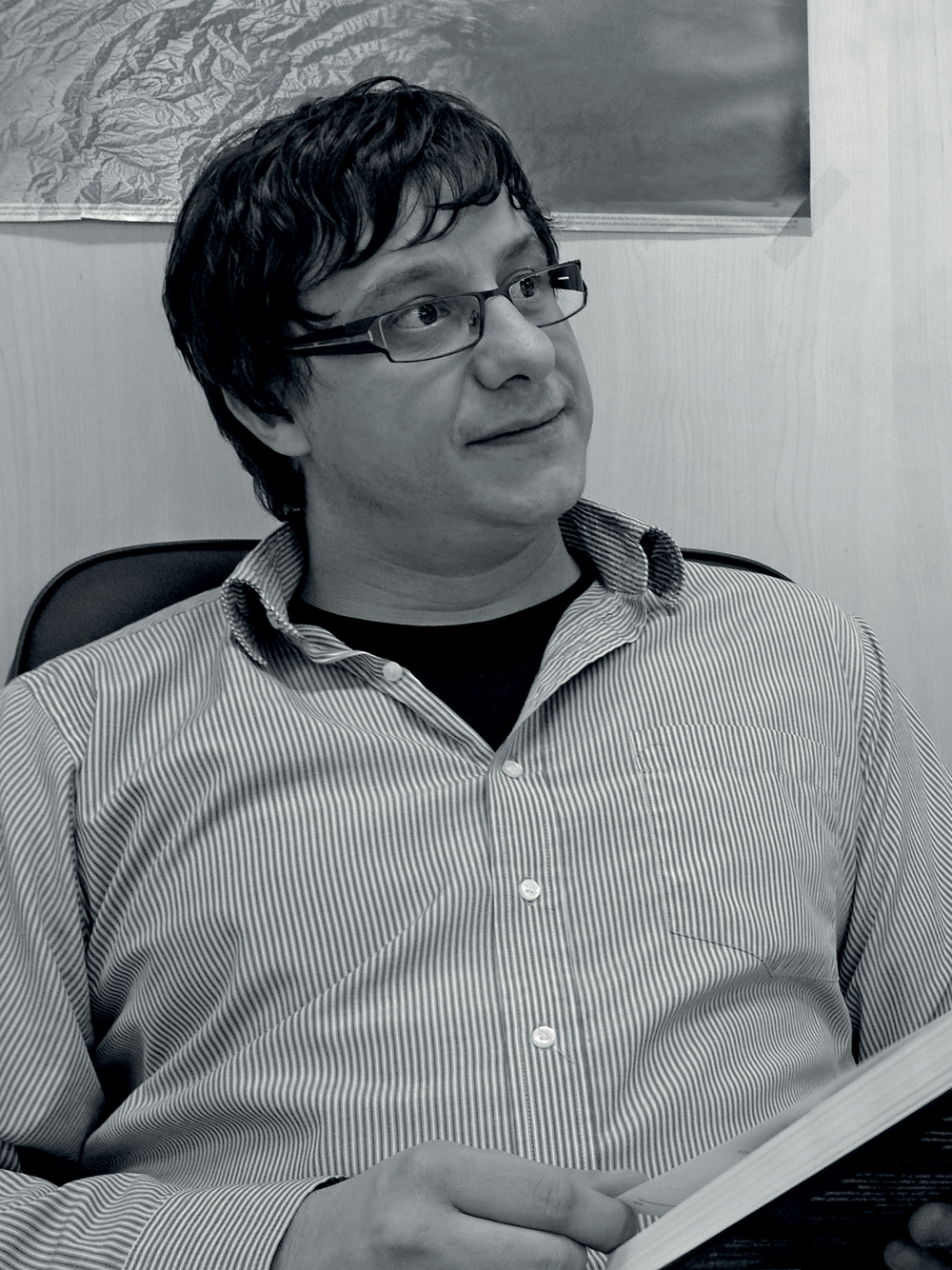}}]{Domen Mongus}
is an Associated Professor at the Faculty of Electrical Engineering and Computer Science, University of Maribor and a member of the Executive Committee of the European Umbrella Organization for Geographic Information (EUROGI). His research interests are Mathematical Morphology, Remote Sensing Data Processing, and Computational Geometry.
\end{IEEEbiography}
\vfill

© 2019 IEEE.  Personal use of this material is permitted.  Permission from IEEE must be obtained for all other uses, in any current or future media, including reprinting/republishing this material for advertising or promotional purposes, creating new collective works, for resale or redistribution to servers or lists, or reuse of any copyrighted component of this work in other works.\\

\end{document}